\documentclass[twocolumn]{aastex631}

\usepackage{url}
\usepackage{natbib}
\usepackage{booktabs}
\usepackage{epstopdf}
\usepackage{graphicx}
\usepackage{amsmath}

\usepackage{array,etoolbox}

\preto\tabular{\setcounter{magicrownumbers}{0}}
\newcounter{magicrownumbers}

\shorttitle{Confinedness of an X3.1 class solar flare}
\shortauthors{Vasantharaju et al}
\begin{document}
	\title{Confinedness of an X3.1 class solar flare occurred in NOAA 12192: Analysis from multi-instruments observations}
	\author{N.~Vasantharaju}
	\email{vasantharaju.naganna@dfa.unict.it}
	\affil{Department of Physics and Astronomy ``Ettore Majorana'', Università degli Studi di Catania, Via S. Sofia 78, I-95123 Catania, Italy}
	\author{F.~Zuccarello}
	\affil{Department of Physics and Astronomy ``Ettore Majorana'', Università degli Studi di Catania, Via S. Sofia 78, I-95123 Catania, Italy}
	\affil{INAF - Catania Astrophysical Observatory, Via S. Sofia 78, I-95123 Catania, Italy}
    \author{F.~Ferrente}
	\affil{Department of Physics and Astronomy ``Ettore Majorana'', Università degli Studi di Catania, Via S. Sofia 78, I-95123 Catania, Italy}
     \author{S.~L.~Guglielmino}
     \affil{INAF - Catania Astrophysical Observatory, Via S. Sofia 78, I-95123 Catania, Italy}
	
\begin{abstract}
The non-association of coronal mass ejections with high energetic flares is sparse. For this reason, the magnetic conditions required for the confinedness of major flares is a topic of active research. Using multi-instrument observations, we investigated the evolution and effects of confinedness in an X3.1 flare, which occurred in active region (AR) 12192. The decrease of net fluxes in the brightening regions, near the footpoints of the multi-sigmoidal AR in photosphere and chromosphere, indicative of flux cancellation favouring tether-cutting reconnection (TCR), is observed using the magnetic field observations of HMI/SDO and SOT/Hinode, respectively. The analysis of spectropolarimetric data obtained by the Interferometric Bidimensional Spectrometer over the brightening regions suggests untwisting of field lines, which further supports TCR. Filaments near polarity inversion line region, resulted from TCR of low lying sheared loops, undergo merging and form an elongated filament. The temperature and density differences between footpoints of the merged filament, revealed by DEM analysis, caused streaming and counter-streaming of plasma flow along the filament and unloads at its footpoints with an average velocity of $\approx$ 40 km s$^{-1}$. This results in decrease of mass of the filament (density decreased by $>50\%$), leading to its rise and expansion outwards. However, due to strong strapping flux, the filament separates itself instead of erupting. Further, the evolution of non-potential parameters describes the characteristics of confinedness of the flare. Our study suggests that the sigmoid-filament system exhibits upward catastrophe due to mass unloading, but gets suppressed by strong confinement of external poloidal field.



\end{abstract}
	
\section{Introduction}
	\label{Intro}
Solar flares and Coronal Mass Ejections (CMEs) are violent explosive phenomena that occur on the Sun. If both these phenomena occur simultaneously and are directed at Earth, they can produce detrimental effects on Earth's magnetosphere and atmosphere. The active regions (ARs) with high magnetic complexity and non-potentiality produce these explosive phenomena \citep{Zirin1987,Schrijver2005}, and when flares are accompanied by CMEs they are referred to as eruptive or else confined/non-eruptive. The association of flares and CMEs has been studied quite extensively and is still an active research topic. Previous findings, for example \citet{Yashiro2006}, showed that the probability of CME-flare association rate increases with the increase in flare strength and the association rate is $90\%-92\%$ for X3.0 class or more intense flares. 

The magnetic flux ropes (MFRs), twisted magnetic field lines wrapped around axial magnetic field, are an essential part of CME structure and support filament/prominence plasma against gravity. There are many possible mechanisms responsible for initiating the outward motion of the MFR and they all come under three main models: ideal magnetohydrodynamic (MHD) instabilities \citep{Torok2003}, flux rope catastrophes \citep{Vantend1978} and magnetic reconnection \citep{Antiochos1999,Moore2001}. One of the popular mechanism in the context of ideal MHD instability is helical kink instability \citep{Torok2004}. Kink instability triggers when the twist of the MFR exceeds the critical twist value of 2.5$\pi$ \citep{Torok2003}. Another relevant mechanism to the present study in the context of MFR catastrophe is ``mass draining'' or ``mass unloading'' effect, which perturbs the equilibrium of MFR. In this mechanism, an upward catastrophe occurs when the mass of the MFR decreases below a critical value \citep{Jenkins2019,Zhang2021}. Under the magnetic reconnection models, the tether-cutting reconnection \citep{Moore2001} between sheared arcades can explain the formation and initiation of MFR eruption successfully. Though these mechanisms efficiently explain the initiation rise motion of the MFR, they failed to explain confined or suppressed eruptions after MFR exceeds the critical values. For example, a statistical study \citep{Jing2018} of  36 strong flare events shows that kink instability plays a minor role in the successful eruption of MFRs. Thus, the rate at which overlying magnetic field decays with height plays an important role in determining the confinedness or successful eruption of MFR. This kind of ideal MHD instability is known as Torus instability \citep{Bateman1978,Kliem2006}. Torus instability (TI) triggers when there is a force imbalance between the outward ``hoop force'' due to the curvature of the MFR and inwardly directed Lorentz force due to the overlying field. It is quantified by a dimensionless parameter, the decay index $n$, which indicates the rate at which overlying field declines with height. MHD simulations provide the onset TI criterion when $n \ge 1.5$ \citep{Torok2005}. In some events, even torus-unstable ($n > 1.5$) flux ropes fail to  erupt and studies were conducted in this direction as well in determining the causes for confinedness of such events. A few notable ones are the dynamic tension force from the external toroidal field \citep{Myers2015}, the Lorentz force due to the non-axisymmetry of the flux rope \citep{Zhong2021} and the rotation of the flux rope \citep{Zhou2019} that all could contribute to the downward Lorentz force in confining the eruption.

\citet{Andrews2003} showed that about 40 \% of M-class flares occurred during the period 1996-1999 are confined and that there are high probability of lack of CMEs association with weaker flares (less than C-class), whereas confined eruptions with more energetic flares are rare. \citet{Schmahl1990} reported about a confined X4-class flare occurred in AR 4492 on 19 May 1984 using radio and X-ray observations. Few more case studies of X-class confined flares are studied by \citet{Feynman1994}, \citet{Green2002}, \citet{Chen2013} and \citet{Liu2014}. \citet{Wang2007} conducted a statistical study of 104 X-class flares during 1996 - 2004 and showed that confined X-class flares, constituting $10\%$ of the sample, occur closer to AR center, while the eruptive flares are at the outskirts. \citet{Cheng2011} performed a comparative study between eruptive (three) and confined flares (six) occurred in AR 10720 and found that eruptive flares have higher decay index in low corona ($< 10$ Mm) than the confined ones.

The AR 12192 is one of the largest, flare prolific and CME poor ARs of solar cycle 24. This AR produced about 35 major non-eruptive flares (29 M-class and 6 X-class) and one eruptive flare (M 4.0) during its disk passage from 18 to 29, October 2014. Many studies were conducted on the X3.1 confined flare event, the strongest amongst the flare series. \citet{Sun2015} and \citet{Sarkar2018} studied the magnetic conditions of the AR and found that the core of the AR exhibits weak non-potentiality, small flare-related field changes and attribute strong overlying magnetic field strength for the confined nature of the flare. \citet{Inoue2016} using nonlinear force free field (NLFFF) extrapolations showed that the core of the AR 12192 is a multi-flux tube system located near the polarity inversion line (PIL) region, where the onset of flare is due to tether-cutting reconnection of low lying field lines of the multi-flux tube system. The confinedness of eruption is attributed to low sheared field lines, which are kink stable, as well as to the strong overlying field strength. \citet{Jiang2016}, using simulations, suggested that an absence of flux rope resulted in confined eruption. On the contrary, \citet{Zhang2017} suggests that that confined flare was due to the complexity of the magnetic field structures.

Past observational and simulation studies could successfully explain the formation of post-flare less sheared core field and stableness against kink instability, but they did not explain the formation of the observed filament and its rise motion during the long duration X3.1 flare event. Owing to the peculiar qualities and rareness of the event, we carried out a comprehensive analysis to investigate the evolution, cause and properties of confinedness of the X3.1 flare using spectropolarimetric imaging data, magnetograms and filtergrams corresponding to different layers of the solar atmosphere obtained by multi-instruments, on board different space-and ground-based telescopes. Description of instruments and data is provided in section~\ref{obs_data}. In section~\ref{ana_res}, we detailed the analysis and results, followed by summary and discussion in section~\ref{summ}.


\begin{figure*}[!ht]
	\centering
	\includegraphics[width=.99\textwidth,clip=]{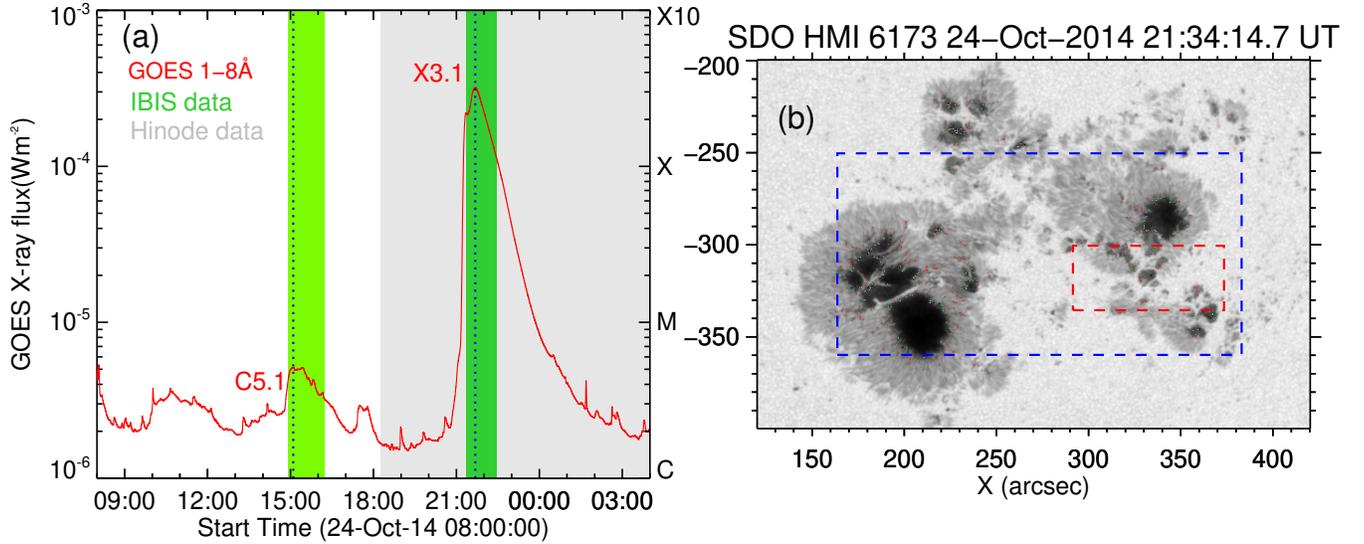}
	\caption{ (a) Disk integrated GOES X-ray flux variations on 24 October 2014. The timeline of data coverage of IBIS and Hinode instruments are shown in shaded regions. (b) HMI/SDO continuum image of NOAA AR 12192 taken near the peak time of the X3.1 class flare. The dashed rectangles in red and blue indicate the FOVs of IBIS and Hinode (BFI/SOT) instruments, respectively. }
	\label{fig1}
\end{figure*}
\section{Observations and Data}
\label{obs_data}
The high spatial, temporal and spectral resolution spectropolarimetric data in the Ca II 8542~$\AA$ line, used in the present analysis, were obtained by the Interferometric Bidimensional Spectropolarimeter (IBIS; \citealp{Cavallini2006,Reardon2008}) at the ground-based Dunn Solar Telescope (DST). The IBIS instrument is based on a dual Fabry–Perot interferometer and mounted in the collimated beam of DST. The Ca II 8542~\AA~line was scanned along 25 wavelength points from 8539.8 to 8544.6 ~\AA, with an average step size of 0.19~\AA~. The pixel size is of $0\farcs095$ and the maximum spatial resolution is about $0\farcs3$. There are two sets of observations available corresponding to two different fields of view (FOVs) over AR 12192, with a total of 144 full spectropolarimetric scans on 24 October 2014. These two sets of observations track portions of  flare ribbons evolution corresponding to two different flares that occurred in AR 12192 on 24 October 2014. The FOV for the first set of observations corresponds to a C5.1 flare (start time - 14:31 UT, peak time - 15:06 UT and end time - 15:54 UT). For this flare, we have IBIS observations from 14:55 UT to 16:42 UT (indicated by the light green shaded region in Figure~\ref{fig1}(a)) and not used in the present work. The FOV for the second set of observations corresponds to the X3.1 flare (start time - 21:07 UT, peak time - 21:41 UT and end time - 22:13 UT). For this flare, we have IBIS observations from 21:20 UT to 22:30 UT. This FOV is marked as red dashed rectangle in Figure~\ref{fig1}(b). Unfortunately, due to poor seeing conditions, we couldn't use the entire data set acquired during this time period. Based on the root mean square (RMS) contrast and visual inspection, we selected only 6 good scans of IBIS data, included in the time interval indicated by the dark green shaded region in Figure~\ref{fig1}(a).

The Solar Optical Telescope (SOT; \citealp{Tsuneta2008,Ichimoto2008}) onboard Hinode, has two filtergraph (FG) instruments called the Broadband Filter Imager (BFI) and the Narrowband Filter Imager (NFI) and a Spectro-Polarimeter (SP). We used filtergrams obtained by the BFI in the Ca II H line (3968 Å) and Stokes-V/I images obtained by NFI in the Na I D1 line (5896 Å). Ca II H and Na I D1 lines are sensitive to the upper and lower chromosphere, respectively. The FOV of SOT was limited to $328''$ x $164 ''$ for the NFI and $218 ''$ x $109 ''$ for the BFI. The blue dashed rectangle in Figure~\ref{fig1}(b) marks the FOV of Hinode filtergraph observations used in the present study. The spatial resolutions of the NFI and BFI are about $0\farcs3$ and $0\farcs2$, respectively. To calibrate Na I D1 V/I data, we used $B_{LOS}$ magnetogram derived from Level 2 dataset of SP.

The H$\alpha$ data acquired from six different telescopes of the Global Oscillation Network Group (GONG; \citealp{Harvey2011}) were used in the analysis of filament evolution during the decay phase of the X3.1 flare. GONG provides full-disk H$\alpha$ images at the cadence of 1 minute, with a pixel size of $1''$.

The Atmospheric Imaging Assembly (AIA; \citealp{Lemen2012}) on board the Solar Dynamics Observatory (SDO; \citealp{Pesnell2012}) produces full-disk Extreme Ultra-Violet (EUV) images in 10 wavelength bands at a high cadence of 12 s with pixel size of $0\farcs6$.  The photospheric magnetic field observations are obtained from the Helioseismic and Magnetic Imager (HMI; \citealp{Schou2012}) on board SDO. Both line-of-sight (LOS) and vector magnetograms (\texttt{hmi.sharp\_cea\_720s} series) obtained at a cadence of 45 and 720 seconds, respectively, are used in this study.

Geostationary Operational Environmental Satellite (GOES) provides the full solar disk integrated soft X-ray (SXR) flux, used to characterize the magnitude, onset, peak and end times of solar flares.

\begin{figure*}[!ht]
	\centering
	\includegraphics[width=.99\textwidth,clip=]{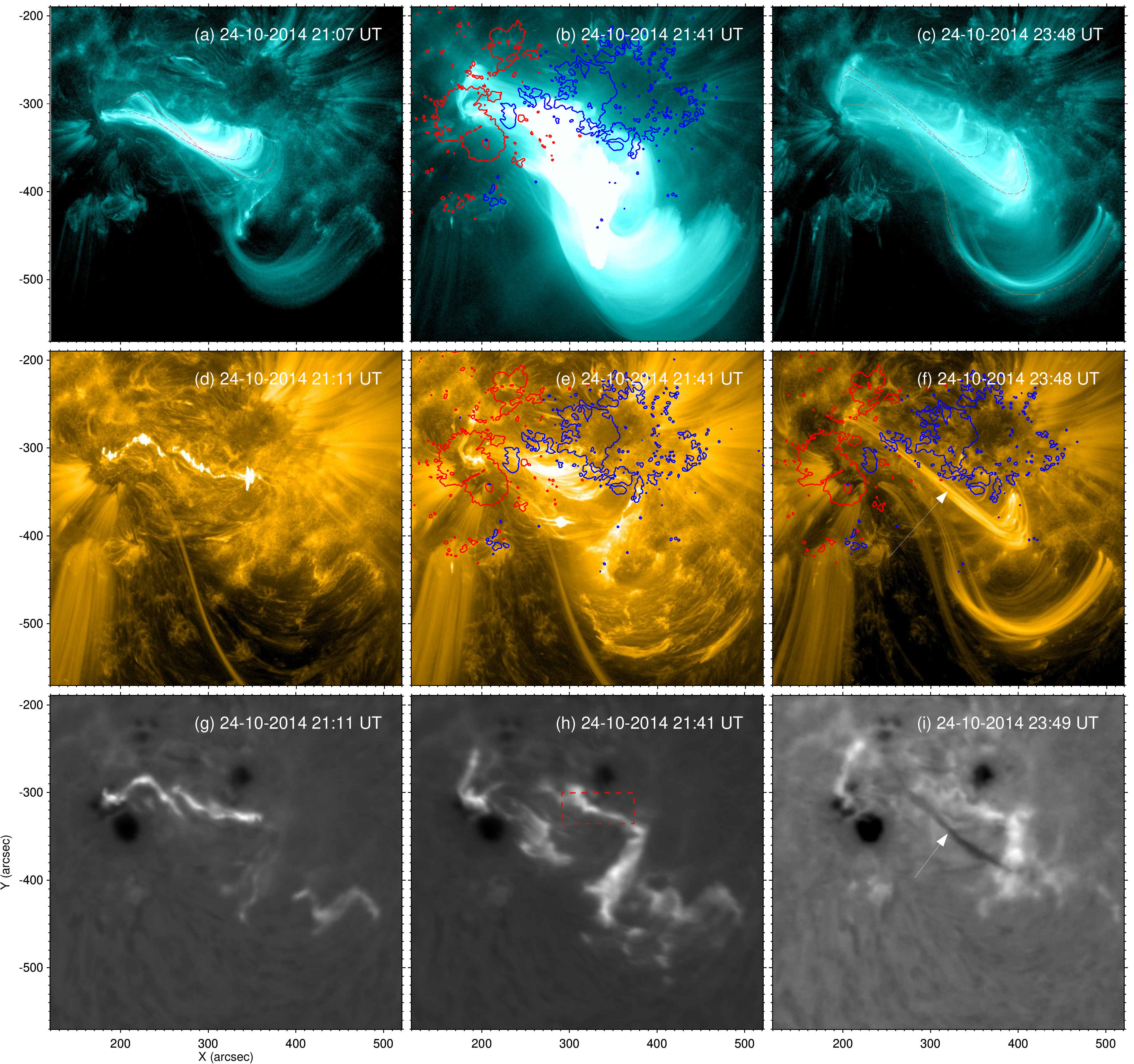}
	\caption{ Images acquired by AIA at 131~\AA, 171~\AA, as well as H$\alpha$ images are used to give an overview of the evolution of AR 12192 during the X3.1 flare. (a)-(c) AIA 131~\AA~images show the morphological changes in the multi-sigmoidal system during the X3.1 flare. (d)-(f) AIA 171 ~\AA~images are at almost same epochs as the top row panels, where HMI $B_{LOS}$ maps with contour levels of $\pm$ 500 G are overlaid. (d) provides evidence of brightenings in the low lying sheared arcade during the onset of flare. (g)-(i) GONG H$\alpha$ images reporting the two flare-ribbon evolution during the flare and the resultant filament formed underneath the sigmoid, as shown in (i). The IBIS FOV is marked by a red dashed rectangle in (h).}
	\label{fig2}
\end{figure*}
\section{Analysis and Results}
\label{ana_res}
\subsection{Overview of the X3.1 flare}
 
The X3.1 flare (SOL20141024T 21:41), the strongest flare produced by AR 12192, occurred at the heliographic location of S16W21. The X3.1 flare was not associated to any CME \citep{Sun2015}, similarly to any other X-class flare produced by this AR. The AR evolved into a highly complex region with a Mount Wilson class of $\beta \gamma \delta$ during its flare prolific period i.e. 20 to 30, October 2014. On 24 October 2014, AR 12192 possessed multi inverse S-shaped sigmoidal loops prior to the X3.1 flare. Images acquired by AIA at 131 and 171~\AA~wavebands are used to represent the morphological evolution of sheared structures during the X3.1 flare in the first and second row panels of Figure~\ref{fig2} (flare prior images are not shown). AR 12192 at the start of the X3.1 flare has multiple sheared structures resembling sigmoids of varying lengths. Two prominent sigmoidal structures are traced by red and blue dashed curves in Figure~\ref{fig2}(a). The brightenings in low temperature channel of AIA 171~\AA~waveband (Fig~\ref{fig2}d) during the onset of the flare indicates that reconnection occurred in between low lying sheared loops rooted at the flare ribbons (Fig~\ref{fig2}g). Figure~\ref{fig2}(b) depicts the peak phase of the flare, flare loops brighten successively from lower to higher atmospheric layers, consequently plasma gets heated up to 10 - 20 MK and an increase in overall brightening is observed. In the decay phase of the flare (Fig~\ref{fig2}c \&~\ref{fig2}f), it is evident that  many pre-flare sigmoidal structures (red and blue dashed curves) are still present and a few more formed as a result of reconnection (orange dashed curve). Due to the flare reconnection, most of these sigmoidal structures hold the filaments underneath and these filaments apparently merge to form a long elongated filamental structure that is shown in Figure~\ref{fig2}(i). The bottom row panels are the H$\alpha$ images obtained by GONG, depicting the chromospheric features evolution during the X3.1 flare at the same epochs as that of top two rows in Figure~\ref{fig2}. Motivated by these observations, we studied the dynamics and non-eruptiveness of filamental structures and the mechanisms responsible for it.


\begin{figure*}[!ht]
	\centering
    \includegraphics[width=8.95cm,height=10.4cm,clip=]{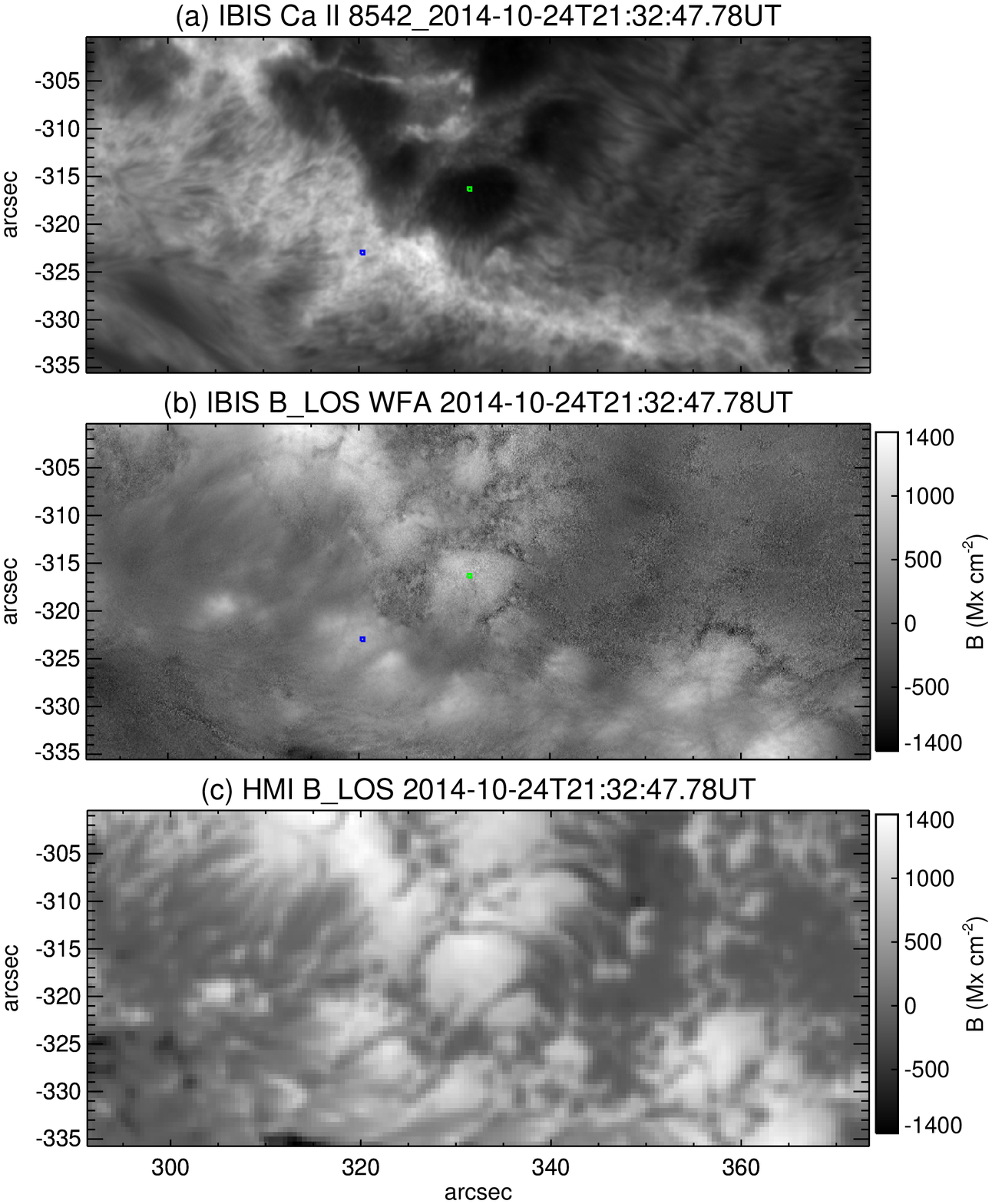}
	\includegraphics[width=8.95cm,height=10.4cm,clip=]{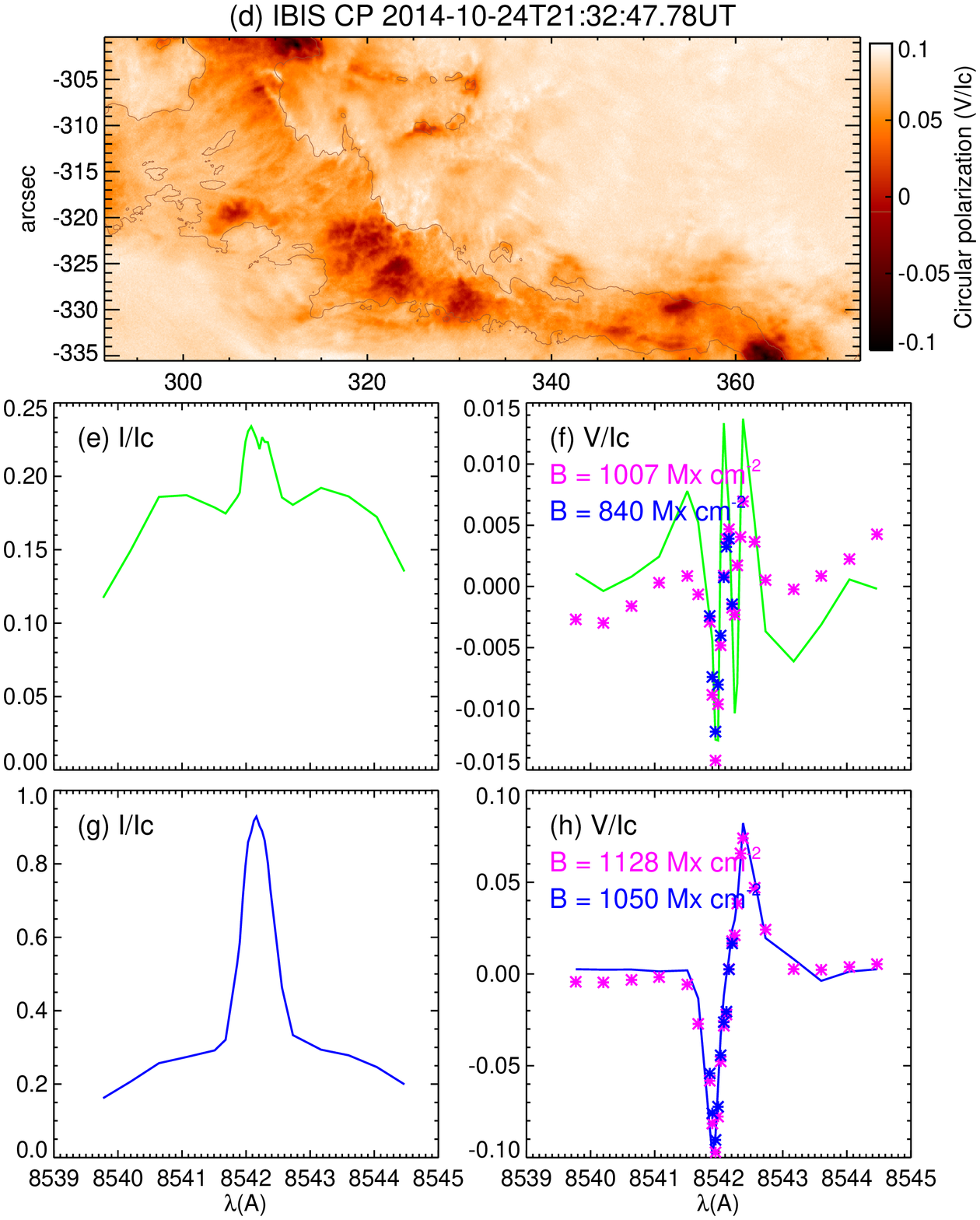} 
	\caption{ Illustration of the results of WFA using a sample scan obtained by IBIS. (a) Stokes I image obtained at the core of Ca II 8542 ~\AA~ line by IBIS. (b) Chromospheric LOS magnetogram deduced by WFA. (c) HMI LOS magnetogram with the same FOV as of IBIS data. (d) Circular polarization (CP) map obtained from IBIS data. CP signals are predominant in the flare ribbon (marked by brown contour), which causes the reconstructed polarity patches within and around the flare ribbon in chromospheric magnetogram (b) have a better match with HMI magnetogram (c). (e) Normalised Stokes I profile of an umbral region indicated by a small green square in (a). (f) V/I profile (solid green) and the two WFA ﬁts obtained from the derivative of Stokes I for the full profile (magenta asterisks) and only for the core proﬁle (Blue asterisks). (g \& h) - Same as (e \& f), but for a small ribbon region (blue square in (a)).}
	\label{fig3}
\end{figure*}
	


\begin{figure*}[!ht]
	\centering
    \includegraphics[width=9.0cm,height=10.4cm,clip=]{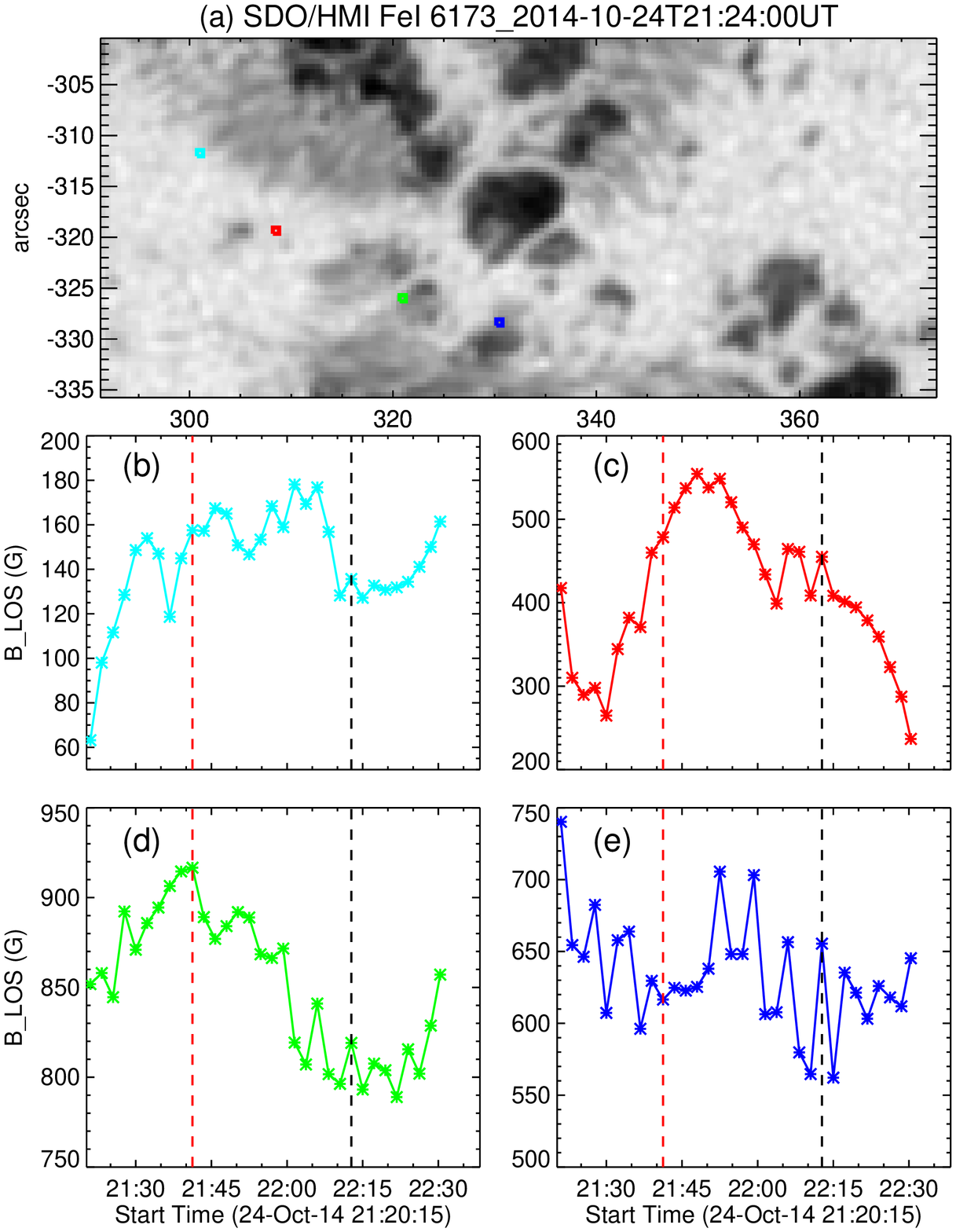}
	\includegraphics[width=8.8cm,height=10.4cm,clip=]{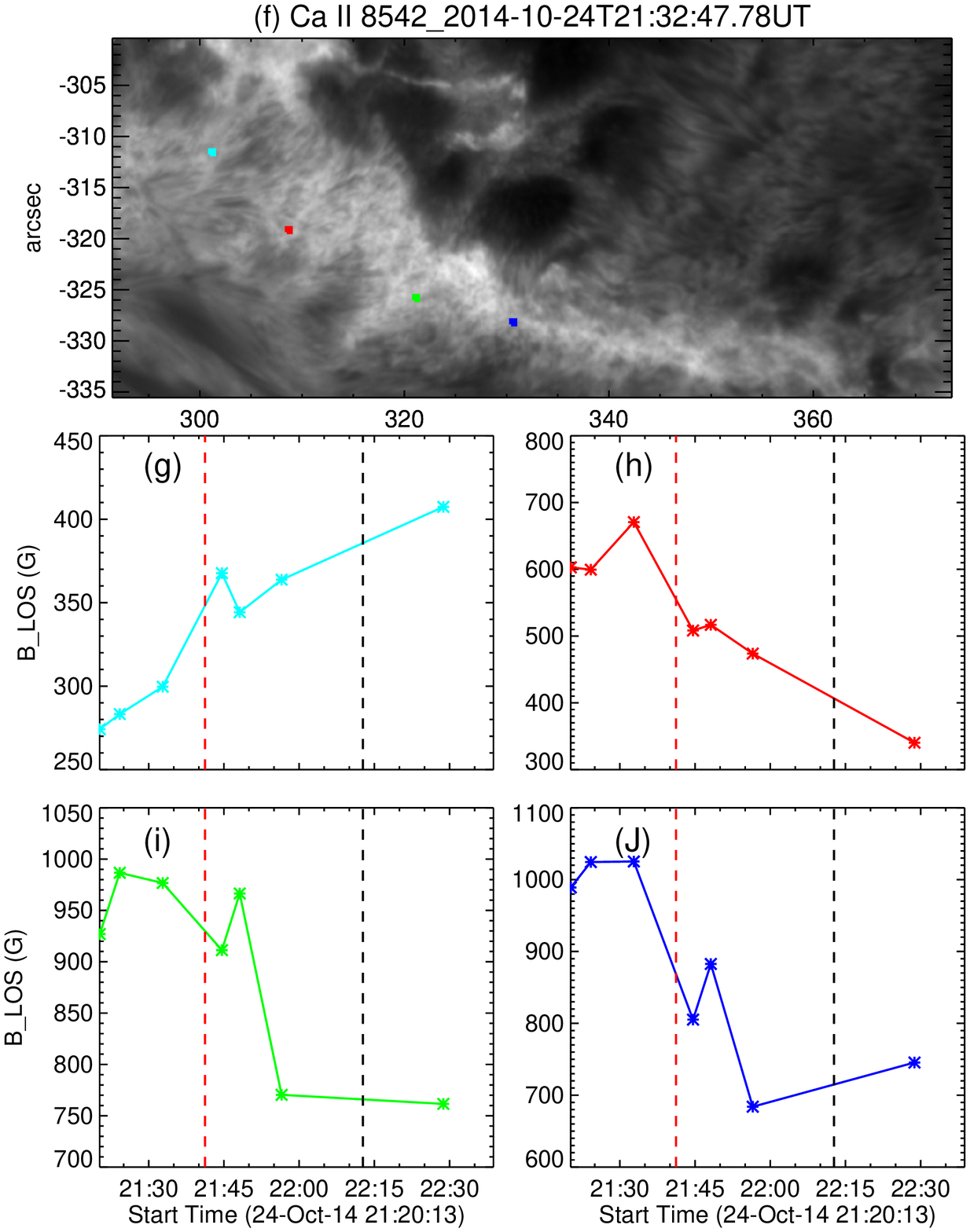}
	\caption{ Comparison of evolution of $B_{LOS}$ values at photospheric (b-e) and chromospheric heights (g-j). (a) HMI continuum image with four different locations in the flare ribbon region marked by four different colored square boxes of 4$\times$4 pixels each. (f) Same as (a) but with Stokes I image of core Ca II 8542~\AA~line. The colored curves in all the remaining panels represent the evolution of $B_{LOS}$ values averaged over the same color of square boxes indicated in (a) \& (f). (h)-(j) $B_{LOS}$ at chromospheric height shows decreasing behaviour indicating untwisting of field lines, while corresponding locations at photospheric height (c-e) shows increasing trend. The dashed vertical red and black lines indicate the GOES peak and end times of the X3.1 flare, respectively.}
	\label{fig4}
\end{figure*}

\subsection{Weak-Field Approximation (WFA) and Changes in $B_{LOS}$}
\label{mag_imp}

Under weak magnetic field limit, the first order perturbation relates the circular polarization profile (V) to the first derivative of the intensity profile (I) with respect to wavelength \citep{Landi1992} :

\begin{equation}
	V_{\lambda} = - \Delta \lambda_{H} ~cos\theta ~ dI(\lambda)/d\lambda
\end{equation}

where the proportionality factor ($\Delta \lambda_{H} ~cos\theta$) depends on the LOS magnetic field, $B_{LOS} = B cos\theta$, with $\theta$ being the angle between the direction of the magnetic field vector and the $B_{LOS}$ component, and the Zeeman splitting is given by:

\begin{equation}
	\Delta \lambda_{H} = (e/4\pi m_{e}c) B {\lambda_{0}}^{2} g_{eff}
\end{equation}

where $g_{eff}$ is the effective Land\'e factor and has the value of 1.1 for Ca II 8542~\AA, B is the magnetic ﬁeld strength, $\lambda_{0}$ is the central wavelength of the spectral line, $e$ is the charge of an electron, $m_{e}$ is the mass of an electron and $c$ is the speed of light.

We determined the chromospheric Line-of-Sight magnetic field $B_{LOS}$ by computing the slope of the linear regression model fit to V and $-dI(\lambda)/d\lambda$ values obtained for each pixel at all the 25 spectral points of the Ca II 8542~\AA~line acquired by IBIS. We applied the WFA separately to the whole line profile (8539.8 - 8544.6~\AA) and the core (8541.8 - 8542.2~\AA) of the line to obtain two values for $B_{LOS}$. The $B_{LOS}$ values obtained from WFA applied to the core of the line profile, indicative of the $B_{LOS}$ values at chromospheric height, are used for further analysis. \citet{Kleint2017} estimated noise in the polarization images to be 1$\%$ of I in 8542~\AA~and considered pixels having V signal strength greater than 2$\%$ of I in deriving the $B_{LOS}$ value using WFA method and found that values less than $\pm$60 Mx cm$^{-2}$ have low S/N. However, we have only considered flare ribbon region for analysis, where V signal has a strength of about 6 - 10\% of I (Fig~\ref{fig3}). Thus, the uncertainty in the derived chromospheric $B_{LOS}$ values should be less than  $\pm$60 Mx cm$^{-2}$.

Figure~\ref{fig3}(a) shows the Stokes I image obtained at the core of Ca II 8542 ~\AA~ line by IBIS at 21:32 UT, while the generated chromospheric ${LOS}$ magnetogram using WFA and the corresponding photospheric ${LOS}$ magnetogram obtained from HMI/SDO are displayed in figure~\ref{fig3}(b) and figure~\ref{fig3}(c), respectively, for qualitative comparison. Though the chromospheric $B_{LOS}$ values obtained from WFA are apparently higher than they are supposed to be, the reconstructed polarity patches are well in agreement with HMI magnetogram.

To illustrate how well the WFA fits with observed profile, we considered two arbitrary pixels located in umbra (green dot) and over flare ribbon (blue dot) as shown in Figure~\ref{fig3}(a)\&(b). WFA fits for the whole line and core of the line profile indicated by magenta and blue asterisks, respectively, are over-plotted on the observed normalized V profile (solid curve) in Figure~\ref{fig3}(f)\&(h). It is evident in Figure~\ref{fig3}(h) that the signal strength of stokes V profile and the fitting of WFA with the observed profile is better than in Figure~\ref{fig3}(f), indicating that the chromospheric $B_{LOS}$ value obtained in the flare ribbon region has a more reliable estimation (less noise) than in umbral region. This is true not just for this particular pixel in flare ribbon, but for all the pixels in the ribbon region, as evidenced in the Circular Polarization (CP) map (Fig~\ref{fig3}d), where CP signals are stronger in flare ribbon region. The mean CP maps were generated by using the equation \citep{Del2003}, $CP = \frac{1}{12<I_c>} \sum_{i=1}^{12} k V_i $, where $<I_c>$ is the average continuum intensity of the quiet sun region within IBIS FOV. In this method, Stokes $V_i$ images obtained at 12 wavelength positions along the Ca II 8542~\AA~line are considered such that 6 wavelength positions are in blue wing and remaining 6 wavelength positions are from red wing of the line. To reconstruct the sign of CP signal, we multiplied Stokes $V_i$ images in blue wing with k=+1 and Stokes $V_i$ images in red wing with k=-1.

As all the twisted field lines in an AR are not related to the flare it produces, \citet{Inoue2016} extensively explored the location of Quasi-Separatrix Layer (QSL;~\citealp{Demoulin1996}) connected to X3.1 flare in AR 12192. QSL is the region of very high magnetic connectivity gradient which favours the formation of thin-current layer, where the magnetic reconnection is considered to occur relatively easy. \citet{Inoue2016} found out that QSLs of X3.1 flare correspond to boundary of flare ribbons. The IBIS FOV, indicated by red dashed rectangles in Figure~\ref{fig1}(a)~\&~\ref{fig2}(h), encloses a part of western flare ribbon which corresponds to the location QSL (refer fig 6 of~\citealp{Inoue2016}).  As the QSLs are the potential sites of magnetic reconnection, the evolution of $B_{LOS}$ in different sub-regions over the flare ribbon during the flare at chromospheric and photospheric heights would be helpful in understanding the orientation and connectivity of field lines.


\begin{figure*}[!ht]
	\centering
	\includegraphics[width=8cm,height=16cm,clip=]{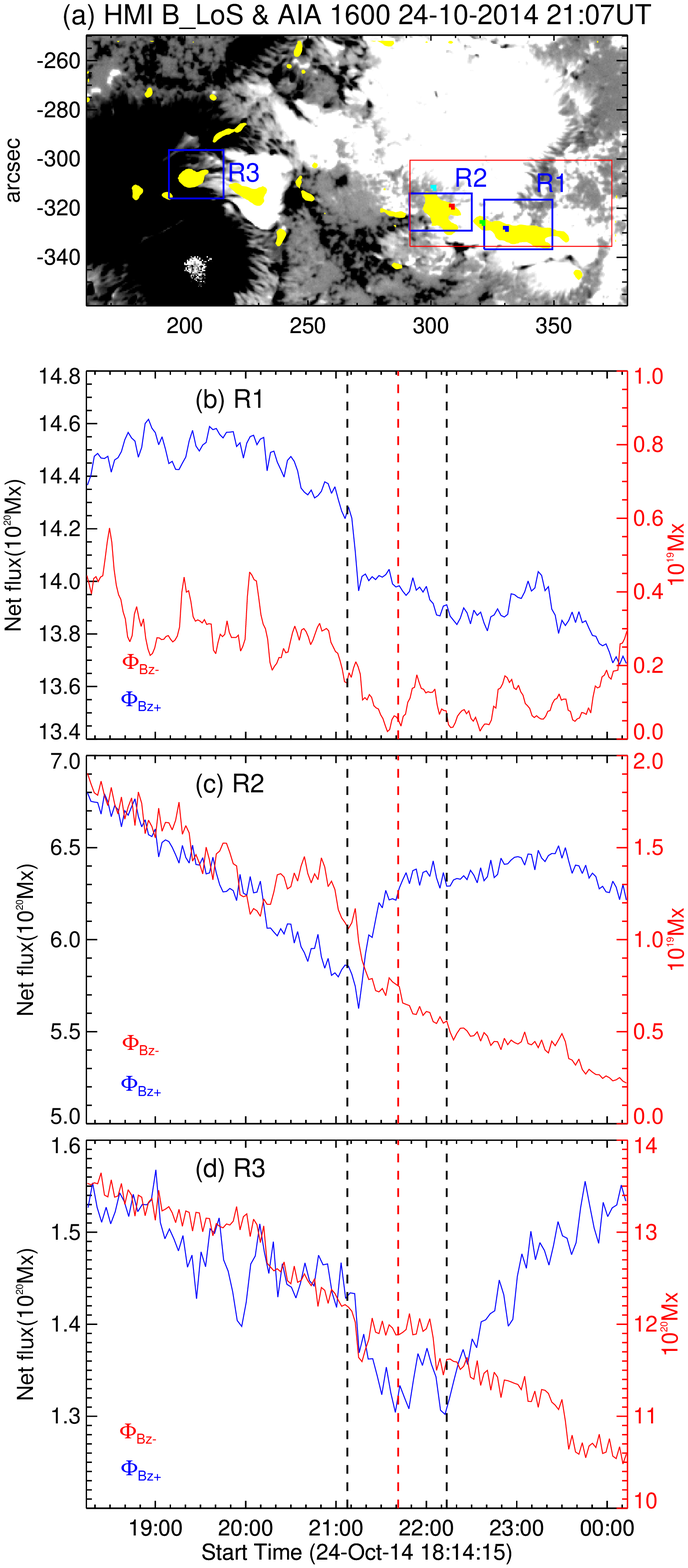}
	\includegraphics[width=8cm,height=16cm,clip=]{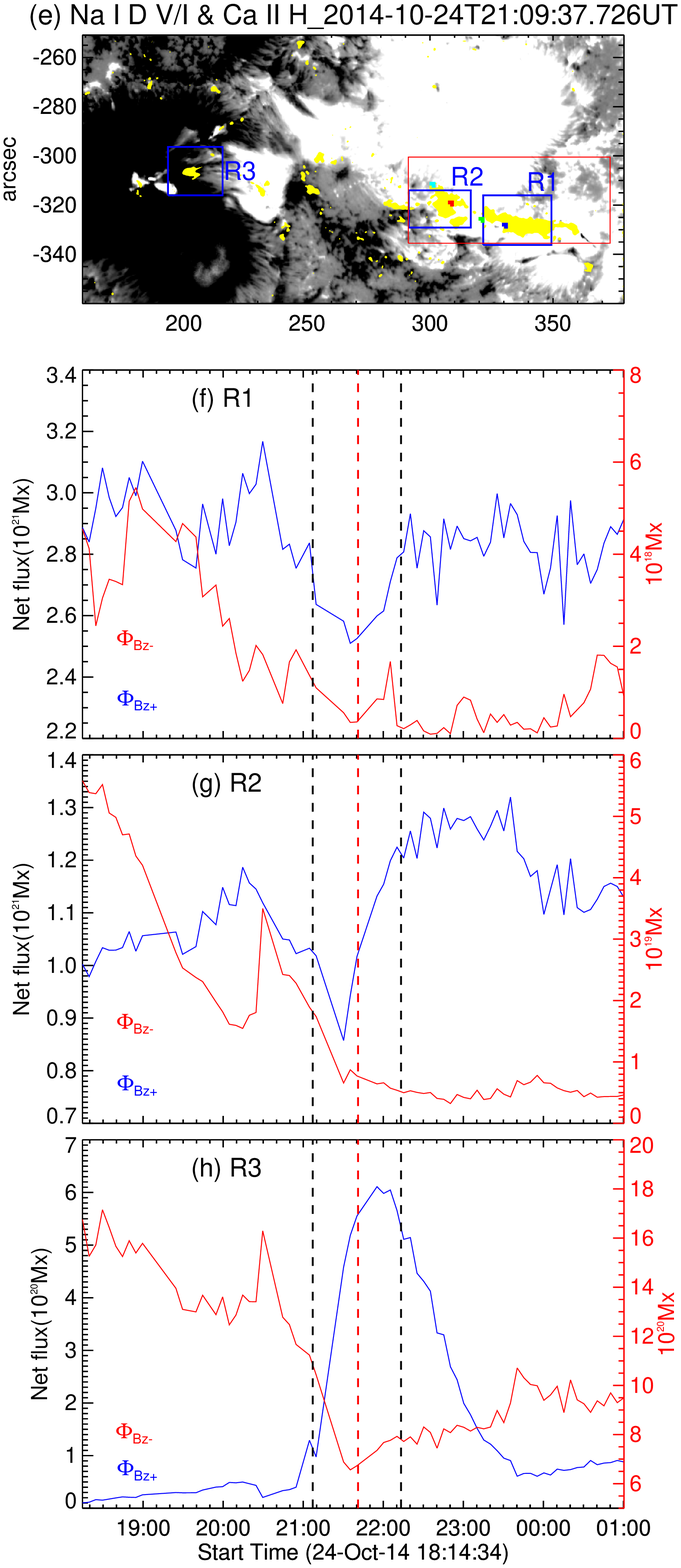}
	\caption{ Magnetic flux evolution near the footpoints of the sigmoidal structure in photosphere (b-d) and chromosphere (f-h) using HMI and Hinode $B_{LOS}$ data respectively. (a) HMI LOS magnetogram overlaid with the yellow color filled contours of initial flare brightening observed in AIA 1600~\AA~waveband. (e) Na I D1 V/I map overlaid with the yellow color filled contours of initial flare brightening observed in Ca II H line.  The decreasing flux content in the positive polarity (blue solid curves) and negative polarity (red solid curves) patches in the three sub-regions signifies flux cancellation. The flare artifacts observed in the chromospheric Na I D1 line camouflaged the decreasing trend of net fluxes (h). The four sub-regions identified in IBIS FOV (Fig~\ref{fig4}a $\&$ f) are also marked in (a) and (e) to specify the location with respect to flare brightening regions. The three dashed vertical lines correspond to GOES start (black), peak (red) and end (black) timings of the X3.1 flare.}
	\label{fig5}
\end{figure*}

The chromospheric $B_{LOS}$, determined from WFA over the IBIS FOV, is studied and compared with the photospheric $B_{LOS}$ obtained from HMI onboard SDO. Though, the strength of chromospheric $B_{LOS}$ determined from WFA over the IBIS FOV is found to be higher than the corresponding strength of $B_{LOS}$ at lower photospheric layer obtained from HMI, the comparison of behavior of temporal variation of $B_{LOS}$ values at these two layers can be studied effectively. Four sub-regions of 4$\times$4 pixels are selected in different locations over the flare ribbon, outlined by squares of different colors in Figure~\ref{fig4}(a)$ \& $(f). The average $B_{LOS}$ values of the 4$\times$4 pixels in four different locations are plotted in four different panels for photosphere and chromosphere separately. We found that $B_{LOS}$ exhibits a decreasing behavior after the flare peak in three sub-regions at both photospheric (Fig~\ref{fig4}c-e) and chromospheric heights (Fig~\ref{fig4}h-j). Conversely, in a sub-region marked by cyan color, $B_{LOS}$ tends to increase (Fig~\ref{fig4}(b) $\&$ Fig~\ref{fig4}g). We would like to note that one of the footpoints of inverse `S' shaped structures (Fig~\ref{fig2}a $\&$~\ref{fig2}d) anchored in the western part of the flare ribbon, are co-spatial with the initial flare brightening regions (yellow filled contours in Fig~\ref{fig5}a $\&$ ~\ref{fig5}e). The decrease of  $B_{LOS}$ in three subregions (lie within initial flare brightening regions) can be attributed to the untwisting of field lines due to magnetic  reconnection, similar to the scenario described in the Figure 8 of \citet{Kleint2017}. The more significant decrease of $B_{LOS}$ is at chromosphere height than at the photosphere is mostly due to the fact that the untwisting of field lines at higher chromospheric height is more prominent than near the footpoints i.e., at the photosphere. The different behavior of $B_{LOS}$ in subregion marked by cyan color from other subregions can be understood more clearly analysing Figure~\ref{fig5}(a)$ \& $(e), where this subregion lies outside of initial flare brightening regions. This indicates that the field lines in this sub-region continue to retain twisted configuration.



\begin{figure*}[!ht]
	\centering
	\includegraphics[width=18cm,height=12.5cm,clip=]{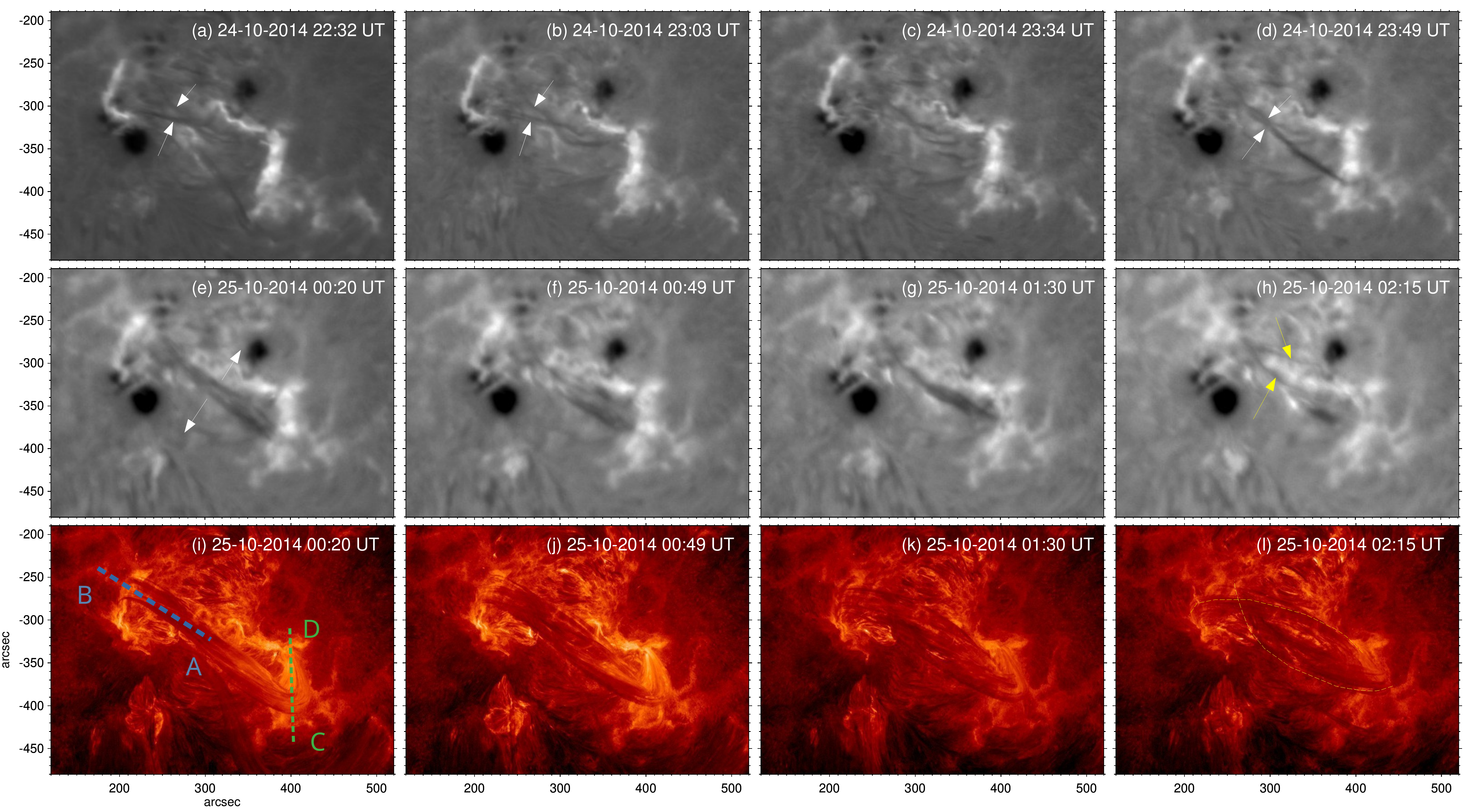}
	\caption{(a-d) GONG H$\alpha$ images showing the merging process of the filaments (underneath sigmoids) near the PIL region. White arrows are used to guide the visualization of the merging process. (e-h) Merged filament splits axially leading to two split filaments indicated by yellow arrows. (i-l) same as (e-h) but with AIA 304~\AA~images, where separated filaments are indicated by traced yellow curves (l). An animation of this figure is available, where AIA 304~\AA~image sequences run from 20:00 UT on 2014 October 24 to 03:00 UT on 2014 October 25 showing the formation of sigmoid-filamental structure along the main PIL, and the subsequent expansion and separation of the structure.}
	\label{fig6}
\end{figure*}

\begin{figure*}[!ht]
	\centering
	\includegraphics[width=18cm,height=13cm,clip=]{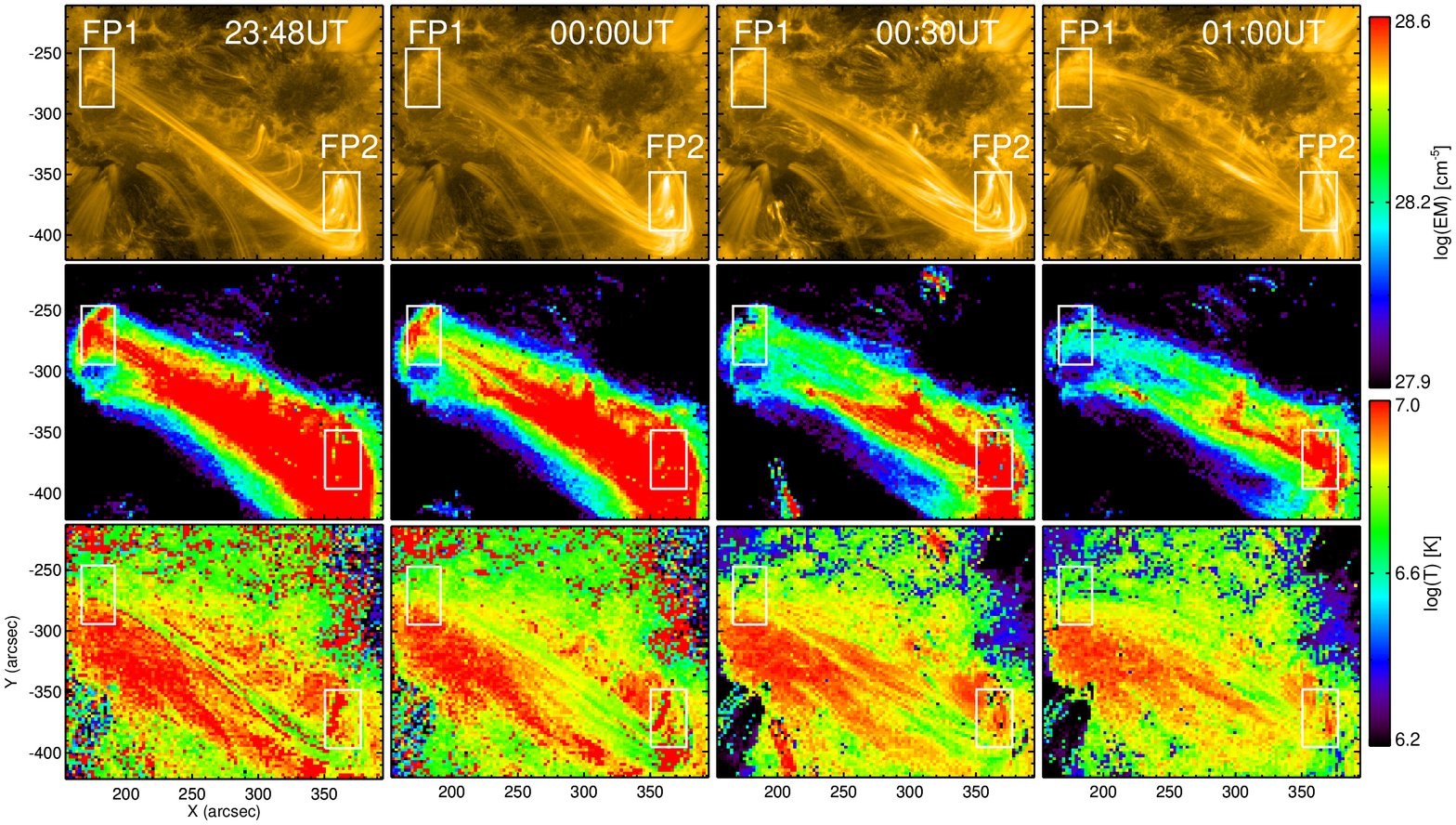}
	\caption{Top row - AIA 171~\AA~ images showing the separation process of the merged sigmoid-filament system. Middle row - Maps of EM distribution at the same epochs as top row panels. Bottom row - $\bar{T}$ maps. Clearly, the distribution of  $\bar{T}$ and EM is higher around sigmoidal footpoint FP2 than near FP1. The two sub-regions (white rectangles) of size $25\times50$ arcsecs are chosen to study the differences in thermal and emission properties between the two sigmoidal footpoints.}
	\label{fig7}
\end{figure*}

\subsection{Flux evolution at Photosphere and Chromosphere}

To compare the temporal evolution of $B_{LOS}$ flux over the initial flare brightening regions in photosphere and chromosphere, we used HMI LOS magnetograms and AIA 1600~\AA~data obtained from SDO with Ca II H and Stokes-V/I of Na I D1 line data obtained from Hinode. The Na I D1 V/I data provide the LOS magnetic field distribution in the lower chromosphere just qualitatively. Na I D1 V/I signal values (dimensionless quantities) are in the range of -1 to +1. To obtain the $B_{LOS}$ quantitative distribution, calibration of V/I data with $B_{LOS}$ data obtained from Hinode Spectro-Polarimeter (SP) of SOT has to be performed \citep{Bamba2013}. We converted the Stokes V/I signal to magnetic field strength in gauss using the regression line equation, $B_{LOS} = 10900 B_{L} -10.21 $, which is derived from the scatter plot of Stokes V/I signals and SP $B_{LOS}$ data obtained before the flare onset (20:30 – 20:50 UT), where $B_{LOS}$ and $B_{L}$  are the converted LOS magnetic field strength in gauss and Stokes V/I values, respectively.

In Figure~\ref{fig5}(a) and (e), HMI LOS magnetogram and calibrated V/I map are overlaid with yellow filled contours of flare brightenings observed during flare onset in AIA 1600~\AA~and Ca II H wavebands, respectively. First, we identified the initial flare brightening regions that are co-spatial with the footpoints of the inverse sigmoidal structure (fig~\ref{fig2}a) and then regions R1, R2 and R3 (blue squares in Fig~\ref{fig5}a $\&$ e) are carefully defined such that they should enclose such initial flare brightening regions at both heights. The flux evolution in these three regions at photospheric and chromospheric heights are shown in Figure~\ref{fig5}(b-d) and Figure~\ref{fig5}(f-h), respectively. At photospheric height (Fig~\ref{fig5}b-d), the decrease of positive and negative flux is clearly observed during the X3.1 flare in all the panels except for region R2, where the positive $B_{LOS}$ flux exhibits increasing trend from the flare start time. This is possibly due to flux emergence in positive polarity of R2. Whereas at chromospheric height (Fig~\ref{fig5}f-h), though the $B_{LOS}$ flux evolution trend appears to be same as that of photospheric height in these regions, flare related artifacts are more prominently visible, especially the sudden increase and decrease of positive flux in region R3 (Fig~\ref{fig5}h). The brightening that appear at R3 region during peak time of flare indicates that the Na I D1 line core at R3 region evidently turned from absorption into emission \citep{Maurya2012}. It is worth to note here that the positive and negative fluxes ranges at two heights are significantly different and these values are obtained from two different instruments. Owing to calibration issues, we can not compare the absolute values of positive and negative fluxes at two heights but their decreasing behavior with time at two heights strongly suggests flux cancellation. The flux cancellation in these brightening regions most likely initiates the tether-cutting reconnection in the sheared arcade which in turn leads to the X3.1 flare. The brightening of sheared loops rooted at the flare ribbons observed in low temperature channel of AIA 171~\AA~(Fig~\ref{fig2}d) indicates that the shorter and lower sheared loops undergo tether-cutting reconnection. Figure~\ref{fig2}(c) indicates that most of the higher sigmoidal structures continue to exist in their sheared form rather than getting relaxed after the flare. Therefore, it is likely that low lying sheared structures are involved in the tether-cutting reconnection, leading to the formation of filaments.

\subsection{Decay phase of the flare}

\subsubsection{Dynamics of the filaments}
AR 12192 holds multi-sigmoidal structures (Fig~\ref{fig2}(a-c)) on 24 October 2014. These multi-sigmoidal structures are observed to carry filaments underneath after the peak phase of the flare and the analysis of the evolution of these filamental structures paves way in understanding the confinedness of the X3.1 flare. The filaments underlying the multi-sigmoidal structures lay one above the other and the evolution of these filaments is displayed in Figure~\ref{fig6} using GONG H$\alpha$ and AIA 304~\AA~images. 

The panels in Figure~\ref{fig6}(a-d) report GONG H$\alpha$ images, showing the merging of filaments (see the white arrows). This process leads to the formation of a merged elongated filament lying over the main polarity inversion line (Figure~\ref{fig6}d). From Figure~\ref{fig6}(e), the merged filament started to undergo separation along its axis. At this epoch, the merged filament started to rise and expand slowly, and during this process it underwent separation. What initiates the rise motion of the merged filament will be discussed in the next subsection~\ref{TE}. In Figure~\ref{fig6}(h), the two distinctly separated filaments indicated by yellow arrows are shown. This separation process of the filament is distinctly visible in AIA $304~\AA$ waveband (Figure~\ref{fig6}(i-l)) as well and separated filaments are marked by yellow dashed curves in (Figure~\ref{fig6}(l)). Based on the visual inspection, it appears that the coronal loops entered into a more relaxed energy state during the process of separation of the filament. 

\begin{figure*}[!ht]
	\centering
	\includegraphics[width=8.7cm,height=13cm,clip=]{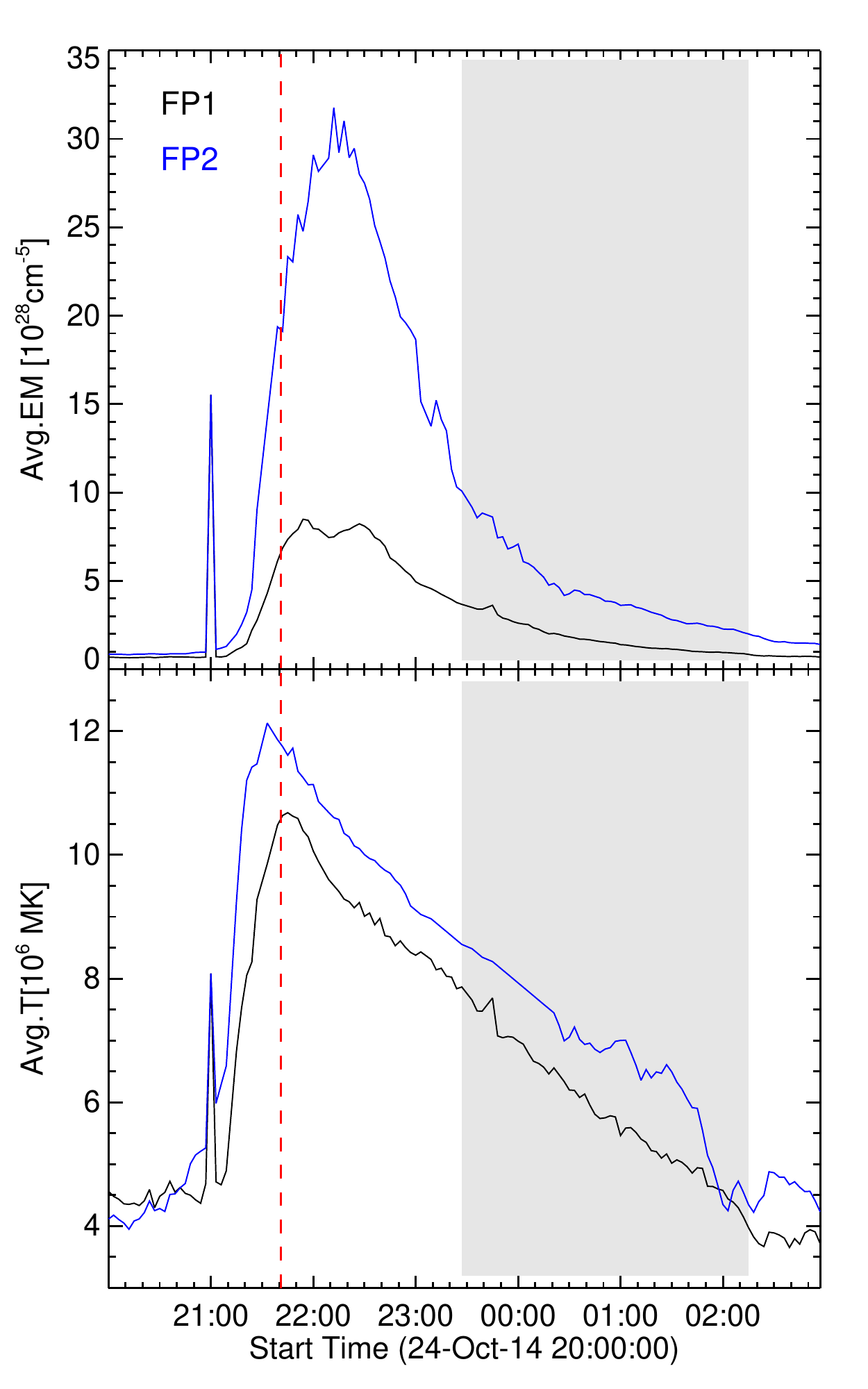}
	\includegraphics[width=9.1cm,height=13cm,clip=]{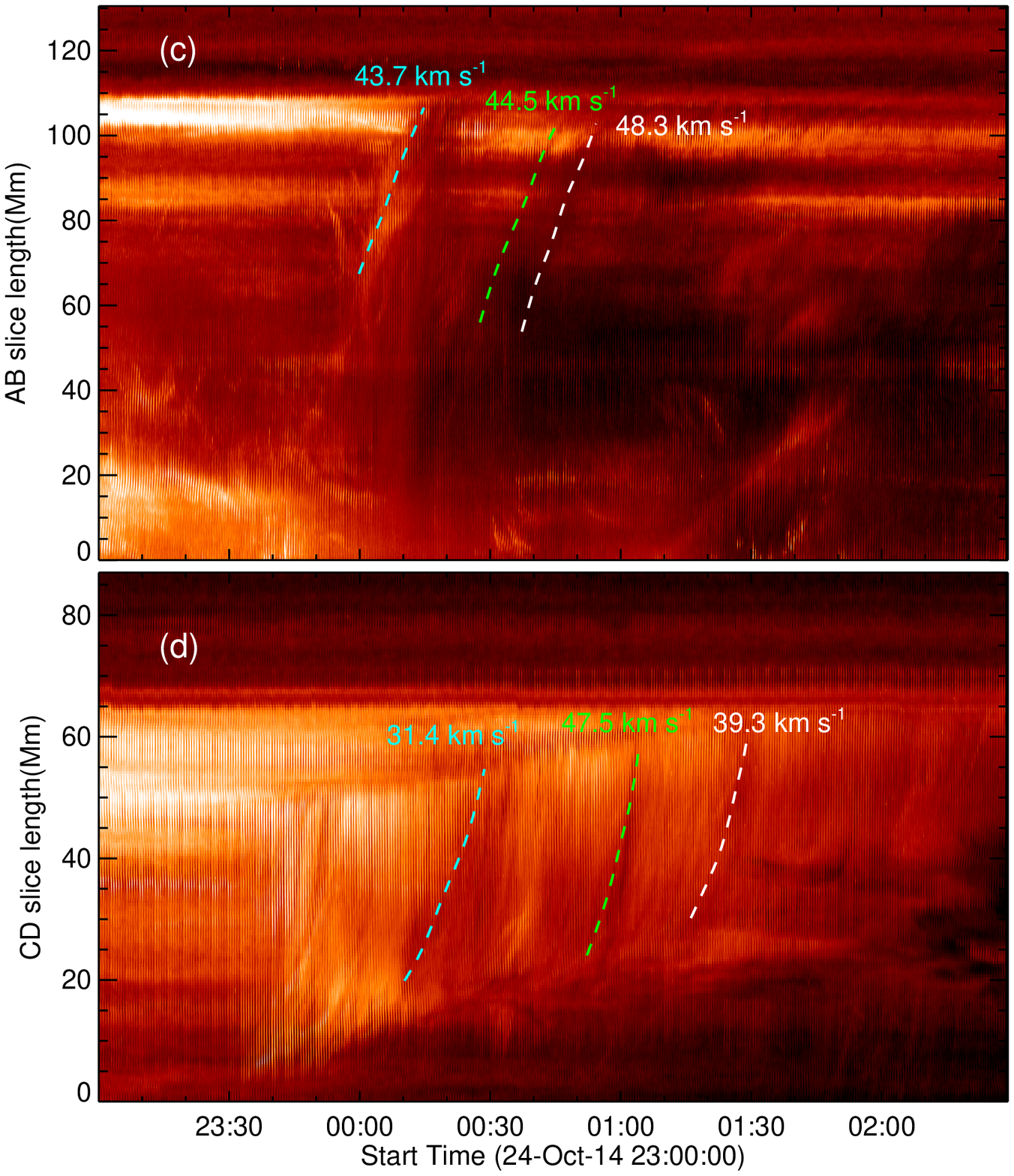}
	\caption{ (a-b) Temporal evolution of average $EM$ and $\bar{T}$ of the regions enclosed by white rectangles, showed in Fig~\ref{fig7}, near the two footpoints of the sigmoidal structure. Solid black and blue curves refer to footpoints FP1 and FP2, respectively. Dashed vertical red line marks the peak time of the X3.1 flare and shaded region indicates the time interval of filament separation. (c) Space-time plot of plasma flow along the slice AB (Figure~\ref{fig6}i) directed towards FP1 and projected flow velocities are annotated against the trajectories of flows in AB. (d) Same as (c) but along the slice CD (Figure~\ref{fig6}i), showing the flow directed towards FP2.}
	\label{fig8}
\end{figure*}

\subsubsection{Emission Measure and Thermal evolution}
\label{TE}
Emission measure (EM) and temperature evolution of the sigmoidal structure holding the merged filament underneath, is studied by applying Differential Emission Measure (DEM) diagnostic technique to six EUV wavebands of AIA/SDO. DEM diagnostic technique allows us to measure the amount of emitting plasma along the LOS with respect to temperature. We used slightly modified version of DEM reconstruction routine \texttt{xrt\_dem\_iterative2.pro} available in Solar Software to work with AIA data, which was initially developed for X-ray Telescope data of Hinode \citep{Golub2004, Weber2004}. Nonetheless, \citet{Cheng2012} applied this code comprehensively on AIA data to study the thermal properties of CME structures. Once DEM(T) maps are reconstructed, the EM and DEM weighted average temperature ($\bar{T}$) can be derived using the following equations: 

\begin{equation}
\begin{split}
    \bar{T} = \int DEM(T) T dt/ \int DEM(T) dt \\
     EM = \int DEM(T) dt
\end{split}
\end{equation}

where integration is performed within the temperature limits of $6.0<LogT<7.1$.

The reappearance of the sigmoidal structure (23:30 UT) in AIA EUV wavebands is co-temporal with the formation of the merged elongated filament in GONG H$\alpha$ observations, as shown in Figure~\ref{fig6}(d) \&~\ref{fig2}(f). DEM analysis is used to understand the rise motion and separation of the merged filament. The maps of $EM$ and $\bar{T}$ in spatial domain are constructed to study the temporal evolution of thermal and emission properties of the sigmoidal structure during the decay phase of the flare i.e., from flare peak time to GOES X-ray flux attaining pre-flare level (24 October 21:41 UT - 25 October 02:15 UT). AIA 171~\AA~images in top row of Figure~\ref{fig7} are used to represent the evolution of the sigmoidal structure, while the corresponding maps of $EM$ and $\bar{T}$ are plotted in middle and bottom row panels, respectively. It is evident from Figure~\ref{fig7} that the two footpoints, indicated by two white rectangles of size $25\times50$ arcsecs, of the sigmoidal structure have different temperature and EM distribution. The footpoint 1 (FP1) region appears to be at lower $\bar{T}$ and EM  distribution than footpoint 2 (FP2) region, which has relatively higher temperature and EM distribution.

To further confirm the asymmetries of these parameters in the two footpoints, we computed the average values of EM and $\bar{T}$ over the two sub-regions enclosing the footpoints of the sigmoidal structure. The temporal evolution of these parameters are shown in Figure~\ref{fig8}(a-b). The time period between the reappearance of the sigmoid/formation of merged filament (24 October 23:40 UT) and the separation of the filament into two distinct filaments until GOES X-ray flux reaches pre-flare level (25 October 02:15 UT) is highlighted in gray shaded region. During this time period, the average EM and $\bar{T}$ values of FP2 (blue curves) are found to be higher than the average EM and $\bar{T}$ values of FP1 (black curves) in Figure~\ref{fig8}(a-b). 

Once the EM distribution is known, the density (n) of the sigmoidal structure can be obtained using $ n = \sqrt{EM/l}$, where l is the width of the sigmoidal structure. As there are no STEREO observations during October 2014, the width of the sigmoidal structure is computed directly on AIA 304~\AA~filtergrams by assuming that the depth of the sigmoidal structure along the line of sight is equal to its width. Before the separation of the filament, i.e. at 23:40 UT, the width of the sigmoidal structure near FP1 and FP2 is estimated to be respectively 8.7 Mm and 10.8 Mm, and the average EM is $4\times10^{28} cm^{-5}$ and $9\times10^{28} cm^{-5}$, corresponding to densities of $6.7\times10^{9} cm^{-3}$ and $9\times10^{9} cm^{-3}$, respectively. Once the filaments get separated i.e. at 2:15 UT, the widths of the sigmoid near FP1 and FP2 increased to 9.9 Mm and 11.7 Mm and the average EM reduced to $0.9\times10^{28} cm^{-5}$ and $2\times10^{28} cm^{-5}$, corresponding to a decreased density of $3.1\times10^{9} cm^{-3}$ and $4.1\times10^{9} cm^{-3}$, respectively. We carried out the similar exercise at the middle of the sigmoid and found that the density decreases from $8.2\times10^{9} cm^{-3}$ down to $2.3\times10^{9} cm^{-3}$. The density of the sigmoid decreases by more than $50\%$ during the process of filament separation. The calculated density values are consistent with past studies \citep{Cheng2012}. These results strongly indicate mass draining or mass unloading from the sigmoid-filament system. 

To compute the velocity of plasma flow along the filament structure, space-time or stack plots were generated using the slits AB and CD as shown in Figure~\ref{fig6}(i). The slits AB and CD are placed on the filament to characterize the trajectories of the plasma flows directed towards footpoints FP1 and FP2, and corresponding space-time plots are displayed in Figure~\ref{fig8}(c) and~\ref{fig8}(d), respectively. Using the trajectories of plasma flows in the stack plots, projected velocities are computed by taking time derivative of smoothed height-time data. It is clear from the stack plots that streaming of plasma along the filament and unloading at its footpoints initiated right after merging of filaments i.e. around 23:45 UT. Initially, the velocity of the plasma flow is slower at FP2 (~$\approx$ 31 km s$^{-1}$) than at FP1 (~$\approx$ 43 km s$^{-1}$); this is most probably due to density differences between the footpoints, where FP2 is at higher density than FP1. Gradually, the flow velocity stabilizes and reaches the sigmoid footpoints with an average velocity of ~$\approx$ 40 km s$^{-1}$, which is consistent with past cases \citep{Wang1999}.

We believe that the temperature and density differences between the two footpoints of the sigmoid-filamental structure lead to streaming and counter-streaming of plasma flow (see animation) within it, which eventually leads to mass unloading at its footpoints. This draining of mass from the sigmoid-filamental structure would reduce the gravitational force acting on it, helping the subsequent ascent and expansion \citep{Low1999}. However, the sigmoid-filamental structure undergoes splitting instead of eruption and the two filament sections start to move apart from one another, as shown in Figure~\ref{fig6}, due to suppression of overlying fields (next section~\ref{np}).

\begin{figure*}[!ht]
	\centering
	\includegraphics[width=.90\textwidth,clip=]{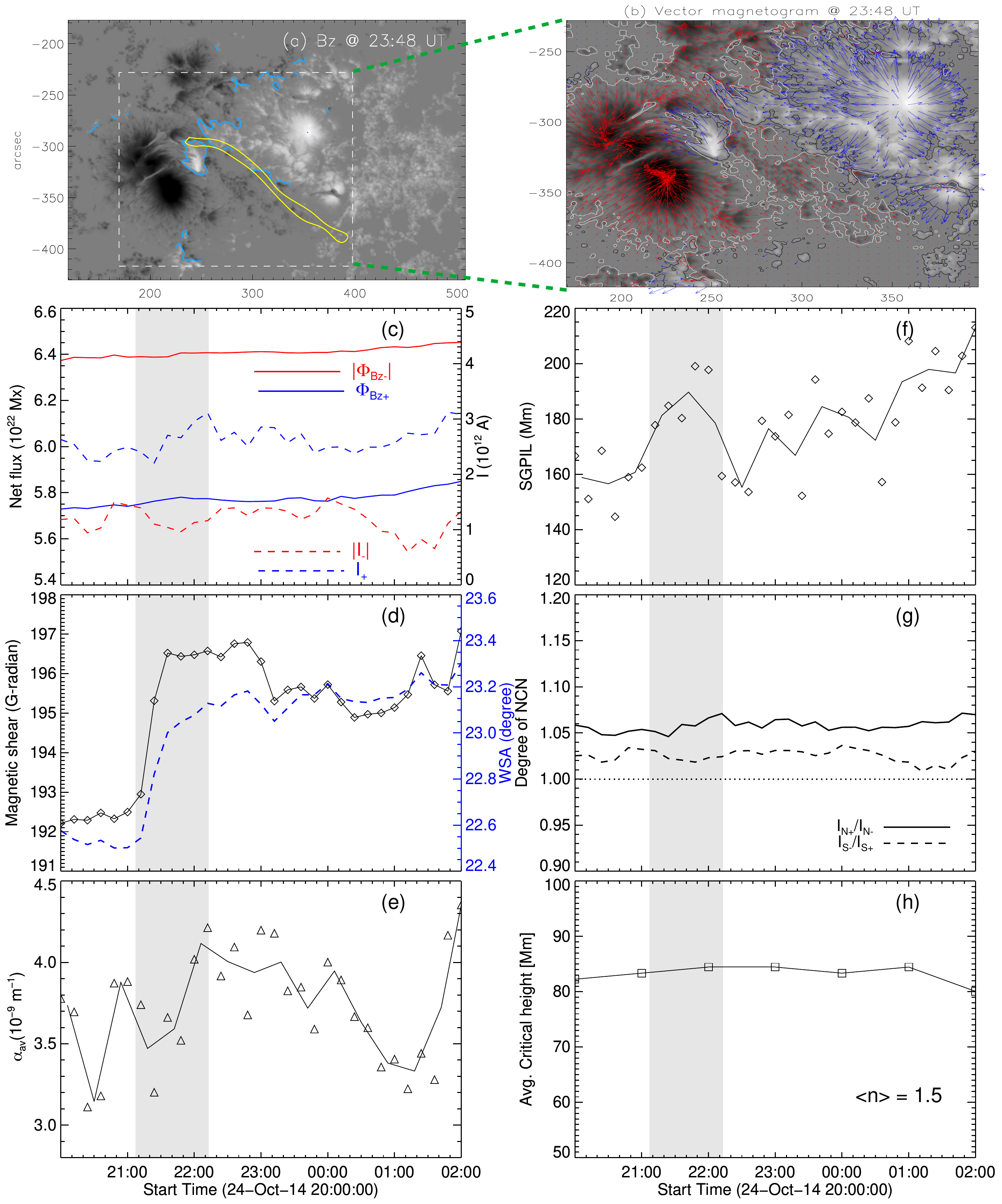}
	\caption{ Temporal evolution of magnetic parameters computed within the region enclosed by the white dashed rectangle shown in (a). Automated Strong Gradient Polarity Inversion Line (SGPIL) traced in blue curves and contour of filament (Fig~\ref{fig6}d) in yellow is overplotted on $B_z$ in (a). (b) Vector magnetogram (transverse vectors overlaid on $B_z$) of the region enclosed by the white dashed rectangle in (a). (c) Net flux and current evolution. (d) Step-wise enhancements of Magnetic shear and WSA. (e) $\alpha_{av}$ evolution (see text) (f) SGPIL (g) Degree of NCN signifying full current neutralization (NCN$\approx$1) and (h) Average critical heights, when $n=n_{c}\ge 1.5$, over the filament channel. The critical heights are computed at every hour throughout 7 hours. The gray shaded region indicates the time duration of the flare as recorded by GOES. Running average (black solid curve) of $\alpha_{av}$ and SGPIL measurements is over-plotted in (f) and (g) to enhance the actual variations.}
	\label{fig9}
\end{figure*}

\subsection{Magnetic non-potentiality and confinedness}
\label{np}
As shown in Figure~\ref{fig2}(c), after the flare the AR continues to hold pre-existing sigmoidal structures and form new sigmoids. This makes the AR quite different from others and therefore we decided to further study the temporal evolution of magnetic non-potential parameters using HMI vector magnetogram data. Basically, the photospheric magnetic non-potential measures are area-dependent, hence we computed the non-potential parameters by taking into account the minimum flux-imbalance condition ($<4\%$) and the maximum field line connectivity involved in flare using AIA EUV images in the flaring area enclosed in the white dashed rectangle in Figure~\ref{fig9}(a). The vector magnetic field map corresponding to the area enclosed by the white dashed rectangle is shown in Figure~\ref{fig9}(b).

The total absolute magnetic flux, given by $\Phi=\Sigma |B_z|dA$ where $B_z$ is the vertical component of the magnetic field, is computed in positive and negative polarity regions and their temporal evolution is observed to be almost constant from 20:00 UT 24 October to 02:00 UT 25 October and is shown in Figure~\ref{fig9}(c). This constant flux in large scale conceals the flux cancellation occurring in small flare brightening sub-regions (Fig~\ref{fig5}) by averaging out the small scale flux cancellation and emergence that occur in small sub-regions. This indicates that the amount of flux decrease due to cancellation process occurring at three small flare brightening regions is not such a significant decrease to reduce the average flux in a large flaring area. During the time interval of 5-6 hours, the total vertical current in positive and negative polarity regions of the flaring area, computed using Ampere's law : $I =\sum (\nabla \times B)_z / {\mu}_o$ also does not show any significant variations (dashed curves in Fig~\ref{fig9}(c)). The total unsigned flux (sum of net fluxes) in the flaring area of the X3.1 flare in AR 12192 is about $1.2\times10^{23}$ Mx, which is in agreement with the recent statistical study of \citet{Li2021}, showing that flares occurring in ARs with the total unsigned flux greater than $1\times10^{23}$ Mx tend to be confined. Thus, the large absolute flux of the AR 12192 could be one of the causes for confinedness of the X3.1 flare.


Magnetic shear is one of the important parameters that account for non-potentiality of magnetic field during flares. Magnetic shear \citep{Wang1994} is defined as the product of observed transverse field strength with the shear angle. Shear angle is the angular separation between directions of observed vector transverse field ($\mathbf{B}_o$) and potential vector transverse field ($\mathbf{B}_p$), and is given by 
$ \Delta\theta = cos^{-1} (\mathbf{B}_o \cdot \mathbf{B}_p / |\mathbf{B}_o \mathbf{B}_p|) $ \citep{Hagyard1986,Ambastha1993}. The Weighted Shear Angle (WSA) is the ratio of summation of magnetic shear to the transverse field strength over all the pixels in the flaring area. It is computed by using $WSA = \sum |B_o| \Delta\theta / \sum |B_o|$. The temporal evolution of the average magnetic shear (solid curve) and WSA (dashed curve in blue) of the flaring area is plotted in Figure~\ref{fig9}(d). The magnetic shear and WSA clearly exhibit rapid, step-wise enhancements during the onset of the flare and continues to remain in a strong sheared state for more than a couple of hours after the flare. The alignment of magnetic transverse field vectors (Figure~\ref{fig9}(b)) are nearly parallel with the main PIL, further substantiating the increase of shear in flaring area in the post flare phase. Again, note that this increase of magnetic shear or WSA during the flare is not due to flux emergence in the flaring area (Fig~\ref{fig9}c). In the past, there have been many studies showing the abrupt and irreversible increase of magnetic shear along the flaring PIL regions during major flares. \citet{Wang1994} showed impulsive  magnetic shear enhancements along the flaring neutral line during six X-class flares and \citet{Wang2012} observed the rapid enhancement of magnetic shear in the localised region of PIL during an X2.2 flare occurred in NOAA 11158. This is mostly caused by the changes in photospheric magnetic fields, especially to the enhancement of horizontal magnetic fields near PIL region \citep{Wang2012,Zuccarello2020,Vasantharaju2022}, as a consequence of coronal implosion during flares \citep{Hudson2008}. However, the result could be different if the analysis is extended from localised regions of the PIL to the whole flaring area. For example, \citet{Li2000} considered the whole flaring area in three ARs to study the changes of average magnetic shear after the flares. They found that magnetic shear in flaring area decreases significantly after the flare. On the contrary, we found that the average magnetic shear in the flaring area of AR 12192 increases after the X3.1 flare. This irreversible increase of magnetic shear is consistent with the fact that no observation of eruption is found. If there were any eruptions, these would have taken away magnetic helicity \citep{Nindos2004}, thereby leading to less sheared post-flare loops. Thus, permanent increase of magnetic shear is an effect of confinedness of the X3.1 flare.

Average $\alpha$ ($\alpha_{av}$) or global $\alpha$ is one of the non-potential parameters used to indicate the degree of twist of magnetic field lines in an AR. It is derived from the z-component of the magnetic field in force-free conditions ($\mu J_z = \alpha B_z $) and can be computed using the equation given by $\alpha_{av}=  \sum [J_z (x, y) B_z(x,y) / |B_z(x,y)|]$ \citep{Pevtsov1994,Hagino2004}, where $B_z$ is the vertical magnetic field and $J_z$ is the vertical current density. The temporal evolution of  $\alpha_{av}$ is plotted in Figure~\ref{fig9}(e). $\alpha_{av}$ exhibits a slight increasing trend during the flare and maintains almost the same value for a couple of hours after the flare, indicating that the twistness of field lines in the AR slightly increases after the flare, which is in support of non-eruptiveness of the flare. Moreover, the magnetic transverse field vector in Figure~\ref{fig9}(b) exhibits a swirling pattern in the upper main negative polarity region: this further corroborates the twistness present in field lines. Thus, non-decreasing $\alpha_{av}$ is also a characteristic effect of confinedness of the X3.1 flare.

Past studies indicate that magnetic gradient is a better proxy than magnetic shear in locating the occurrence and productivity of flares and their strength in an AR \citep{Wang2006,Vasantharaju2018}. We therefore computed the Strong Gradient PIL (SGPIL) using an automated procedure described in \citet{Vasantharaju2018}. In this procedure, vertical gradient maps and potential field are computed using a smoothed $B_z$ map and applying the threshold of potential transverse field  ($>300$G) and strong magnetic field gradient ($>50$ G/Mm) to the zero gauss contours on smoothed $B_z$ maps. The SGPIL length evolution in time is plotted in Figure~\ref{fig9}(f). Total SGPIL length decreases from 185 Mm (peak flare) to 155 Mm after the flare, only for a short time interval. Thereafter, strong gradients near PIL start to increase so as the SGPIL length. SGPIL segments are fragmented, scattered and not continuous in the flaring area of AR 12192. Mostly, the twisted flux rope resides above the continuous high gradients of PIL region, whereas the flaring area in AR 12192 possesses fragmented and scattered SGPILs which might indicate the AR's inability to host strong, long and continuous flux-rope capable of eruption \citep{Vemareddy2019}. Thus, fragmented SGPILs in the AR could also be one of the causes for confinedness of the X3.1 flare.

\citet{Liu2017} suggested that the degree of Net Current Neutralization (NCN) would be a better proxy than strong shear or gradients near PIL in assessing the eruptive nature of flares from an AR. The net current in each polarity has both positive and negative components. The NCN is computed for each polarity by obtaining the ratio of Direct Current (DC) and Return Current (RC) \citep{Torok2014}. DC and  RC  are computed by integrating vertical current density values of different signs separately. The temporal evolution of  $|DC/RC|$ values in both polarity regions are plotted in Figure~\ref{fig9}(g). The DC is found to be positive in the north polarity and negative in the south polarity. The $|DC/RC|$ values in both polarity regions are almost equal to unity. Past studies \citep{Liu2017,Vemareddy2019} showed that the full current neutralization (NCN=1) is a characteristic of a non-eruptive AR, indicating the absence of direct-current channels over the PIL region, whereas an AR characterized by non-neutralization (NCN $>$ 1.3) of currents is prone to erupt. Thus, the full current neutralization in AR 12192 for an extended time interval leads to produce many confined flares, including the X3.1 flare under study.

\begin{figure*}[!ht]
	\centering
	\includegraphics[width=5.7cm,height=5cm,clip=]{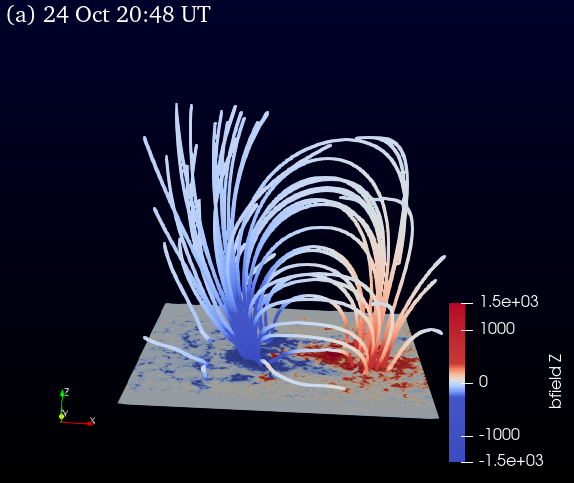}
	\includegraphics[width=5.7cm,height=5cm,clip=]{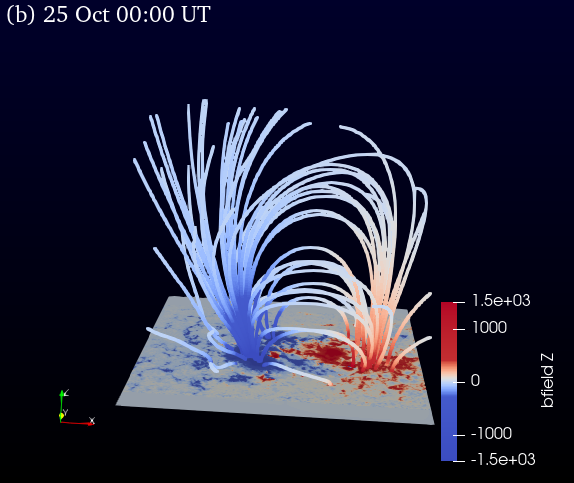}
	\includegraphics[width=5.7cm,height=5cm,clip=]{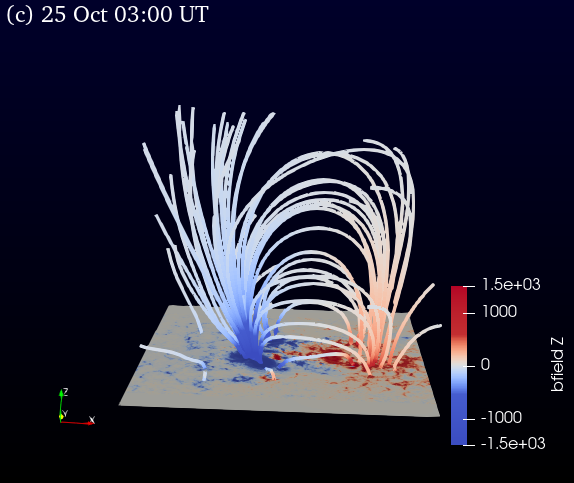}
	\includegraphics[width=5.7cm,height=5cm,clip=]{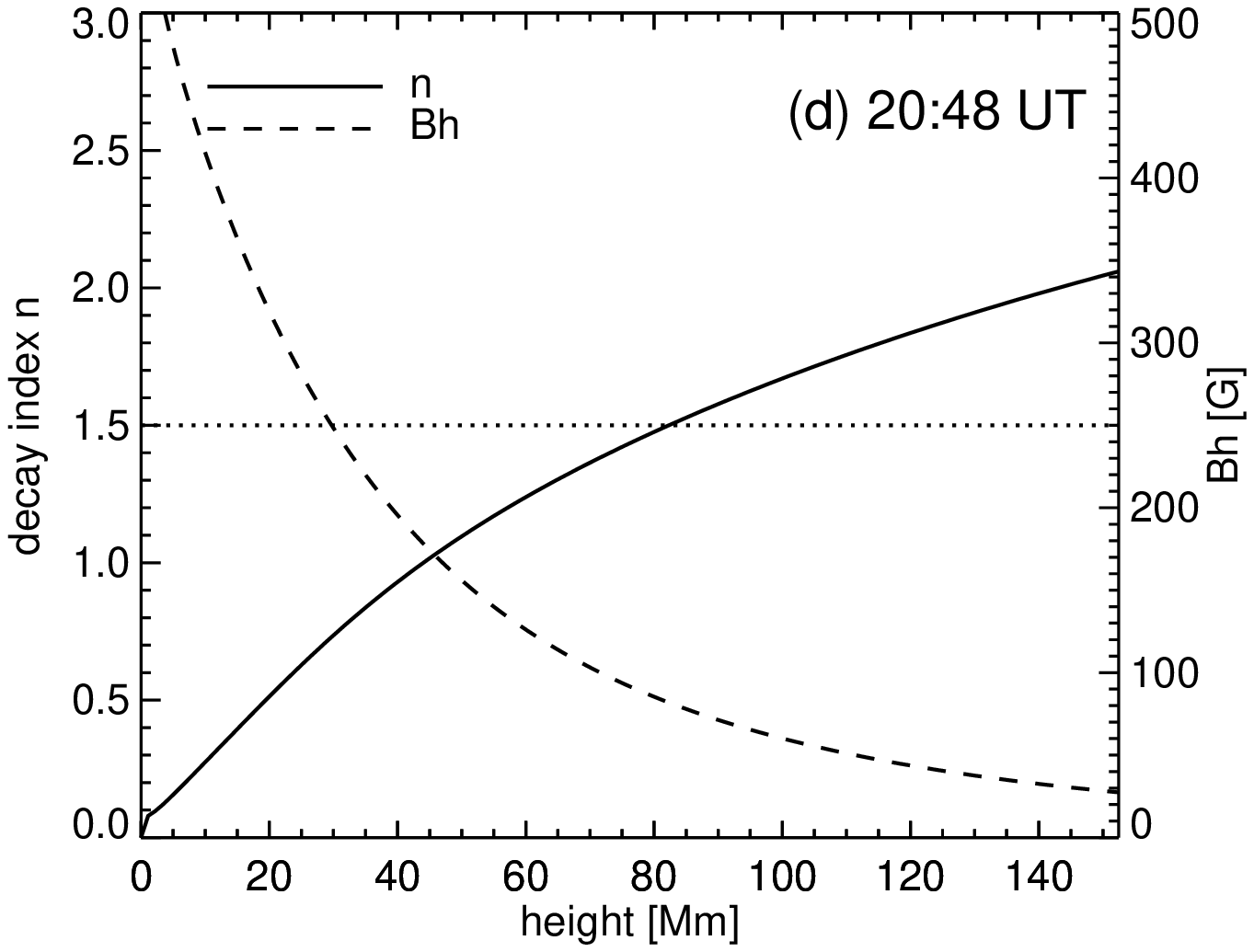}
	\includegraphics[width=5.7cm,height=5cm,clip=]{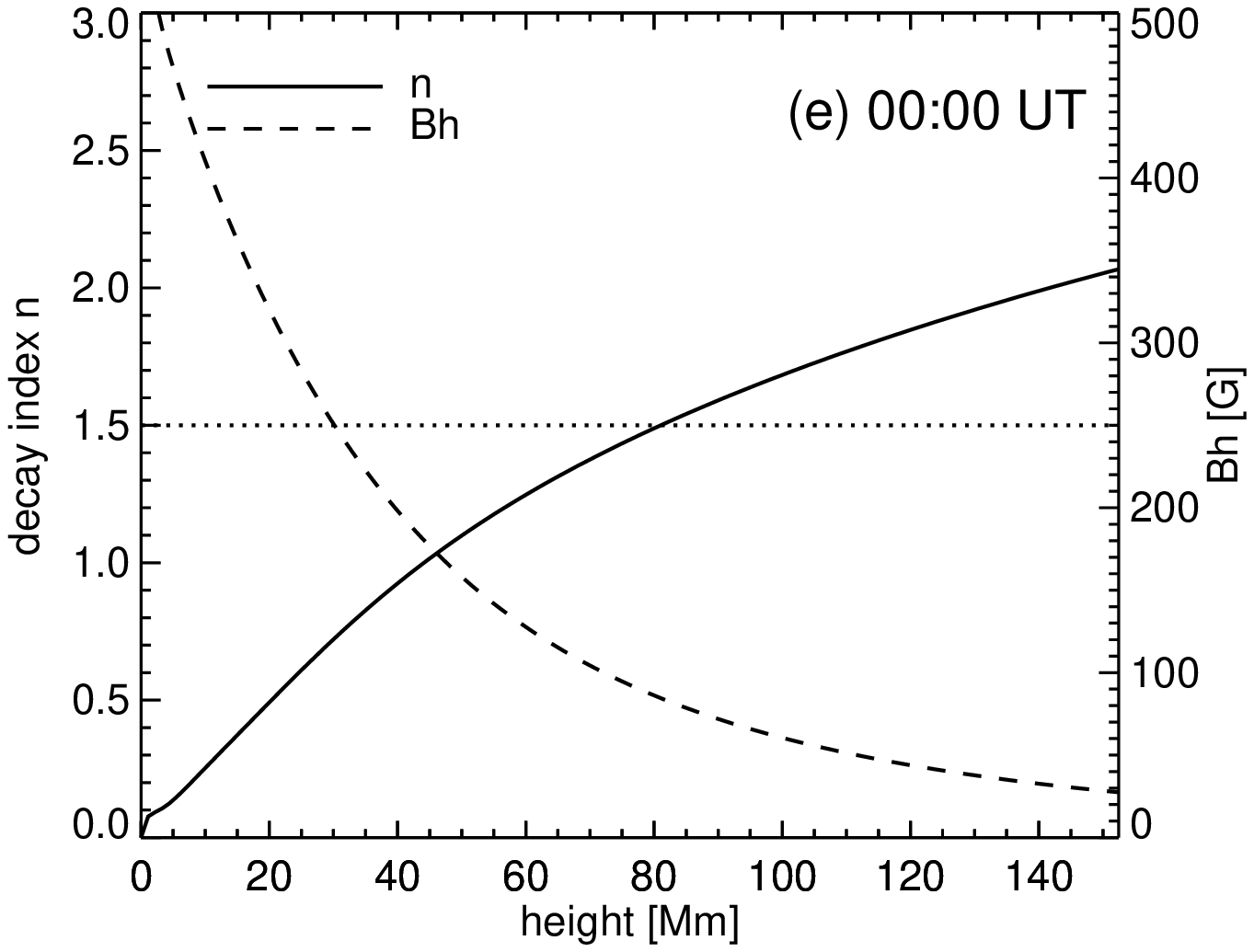}
	\includegraphics[width=5.7cm,height=5cm,clip=]{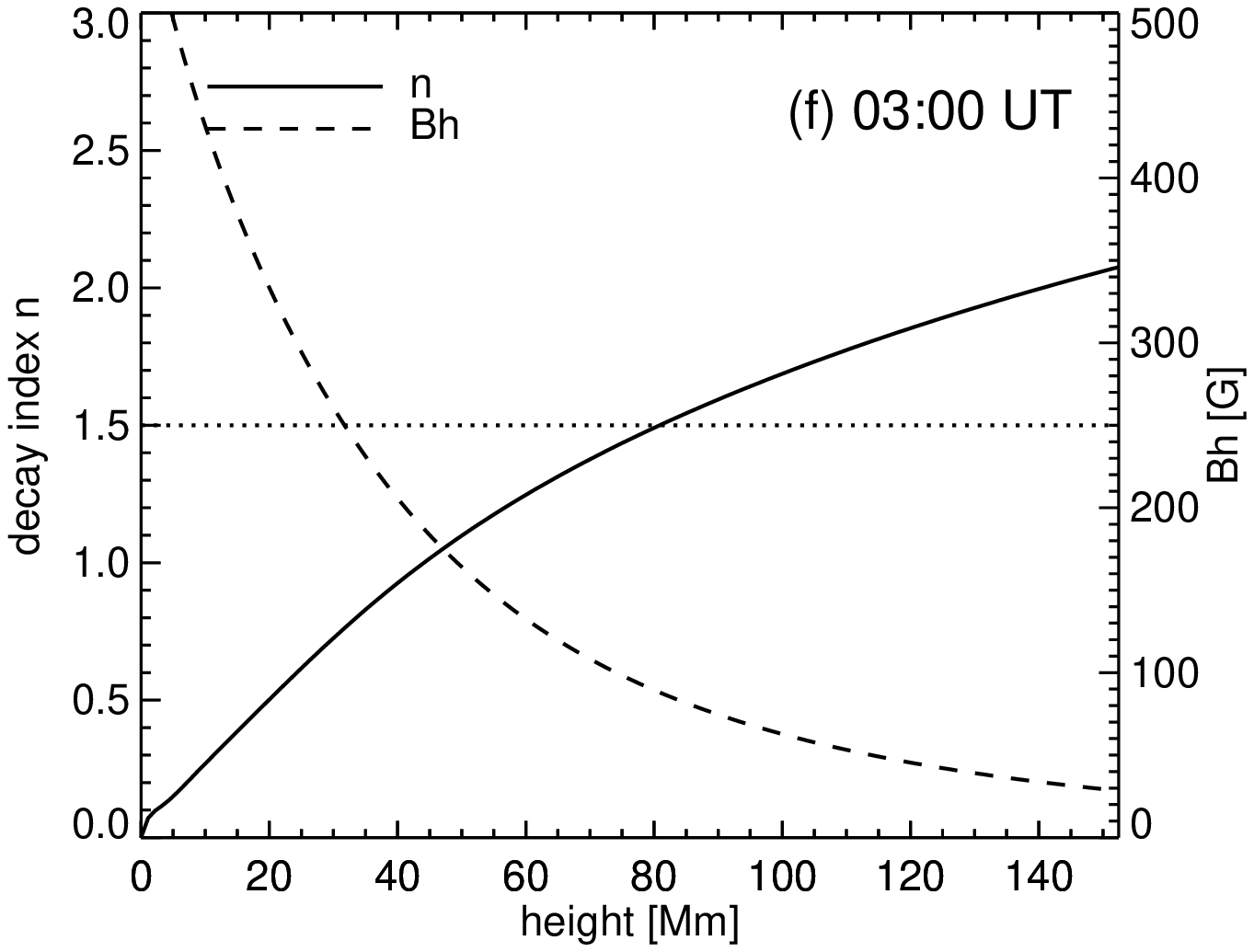}
	\caption{(a-c) PFSS configuration of AR 12192 at different time instances, from pre-flare time (20:48 UT) to end time (3:00 UT), of X3.1 flare. The unvarying potential field configuration provides the robust confinement throughout the flare. (d-f) Corresponding variations of decay index and horizontal magnetic field strength with height. Decay index, n, reaches 1.5 at about 80 Mm above the photosphere.}
	\label{fig10}
\end{figure*}

It's worth noting that the distribution of fragmented SGPIL in the flaring area and the full current neutralization (NCN=1), both indicating absence of robust flux rope along PIL, may contribute to the confinedness of the X3.1 flare. However, we observed the appearance of sigmoid-filament structure along the main PIL and its dynamics of rise and expansion. Thus, the main contribution to the confinedness of X3.1 flare should be the stronger inward directed force from background field and not the weaker outward driving force from the inner non-potential magnetic field. So, we examined the role of background coronal magnetic field using the Potential Field Solar Surface (PFSS; \citealp{Schrijver2003}) approximation. The lower-boundary data is provided by HMI vertical component of the magnetic ﬁeld (SHARP series). The decay index is defined as, $n(z) = -\frac{z}{B_h} \frac{\partial B_h}{\partial z}$, where $z$ is the geometrical height from the photospheric surface and $B_h$ is the horizontal field strength. After the coronal field extrapolation, $B_h$ as a function of $z$ along the filament (yellow contour region in Fig~\ref{fig9}a) is obtained. We repeated the procedure on a time interval of one hour from 20:00 UT on 24 October to 02:00 UT on 25 October. We then plotted the average decay index along the filament channel and $B_h$ as a function of $z$ at each hour (Fig~\ref{fig10}d-f) and we found that the decay index reaches the theoretical critical value of 1.5 \citep{Torok2005} beyond 80 Mm above the surface. The temporal evolution of average critical height above the filament channel (or main PIL) is plotted in Figure~\ref{fig9}(h). Past statistical studies like \citet{Vasantharaju2018,Baumgartner2018,Li2020} showed that the ARs producing confined flares mostly tend to have high critical heights above 50 Mm, owing to strong confinement, whereas for eruptive flares, critical heights are less than 42 Mm, indicative of weaker overlying field strength. For AR 12192 during the X3.1 flare, almost constant critical height of about 80 Mm throughout the flare duration of 6-7 hours, indicates that the background magnetic field strength is strong enough to confine any possible eruption. 

Furthermore, \citet{Myers2015} using laboratory experiment showed that the orientation of external potential field configuration  with respect to flux rope axis is necessary to determine the specific component of downward Lorentz force. The total potential magnetic field is the superposition of strapping field, running perpendicular to the flux rope axis and guide field, running toroidally along the flux rope axis. The coronal field lines rendering extrapolated using PFSS approximation at different time instants throughout the flare duration is shown in Figure~\ref{fig10}. Figure~\ref{fig10}(a-c) displays the potential field configuration of AR 12192 at different stages of the X3.1 flare and panels (d-f) show the corresponding  variations of decay index and horizontal magnetic field strength with height. The filament axis is lies along the PIL of two main polarities (yellow contour in Fig~\ref{fig9}a). From the PFSS plots, it appears that the direction of external poloidal magnetic field is oriented nearly perpendicular to the axial direction of filament. This indicates that the strapping force is more dominant  downward force than the dynamic tension force, induced by the toroidal field. Thus, we opine that the main contributor to the downward Lorentz force towards confining the X3.1 flare would be the strong strapping field.

\section{Summary and Discussion}  
\label{summ}
In this paper, we investigated the nature of confinedness of an X3.1 flare originated in AR 12192 in different layers of the solar atmosphere using the multi-wavelength observations obtained from ground- (IBIS \& GONG) and space-based (Hinode \& SDO) instruments. The X3.1 flare (strongest among the flares produced by AR 12192) was of long duration, lasting for 5 - 6 hours and occurred at a heliocentric angle of $\mu = 0.9$. The AR holds multi sigmoidal structures prior to the start of the flare. Low lying sheared field lines underwent tether-cutting reconnection during the flare, bringing minimum morphological changes to the high lying pre-flare coronal sigmoidal structures, but showing the appearance of filaments underneath these sigmoids. These sigmoid-filament systems lying one over the other exhibit dynamic behavior of merging and subsequent separation. The temperature and density differences between the footpoints of the merged sigmoid-filament system, as revealed by DEM analysis, aids in understanding the separation and non-eruptiveness of the merged filament. Confinedness of the X3.1 flare is mainly caused by the strong confinement provided by the external magnetic field rather than the weaker non-potentiality of the core AR. 
 
AR 12192, being located in the southern hemisphere, shows positive helicity and follow the dominant helicity sign rule \citep{Pevtsov1995}, but it shows inverse S-shaped sigmoids on 24 October. Generally, inverse S-shaped sigmoids predominantly appear in northern hemisphere \citep{Rust1996}, which makes the AR 12192 unconventionally twisted. EUV/AIA observations reveal that the AR has multi sigmoidal structure. Moreover, brightening of the flare loops with their footpoints rooted at flare ribbons observed in low temperature channel of AIA 171 A (Fig~\ref{fig2}d) provides evidence that the shorter and lower sigmoidal loops undergo magnetic reconnection. The $B_{LOS}$ flux cancellation at both photospheric and chromospheric heights in the brightening regions, which are co-spatial with footpoints of low lying sheared field lines, supports the idea of tether-cutting reconnection \citep{Moore2001} to produce the X3.1 flare and is in agreement with the numerical study of \citet{Inoue2016}. Further support of tether-cutting reconnection comes from the analysis of a part of the flare ribbon area (segment identified as QSL in \citealp{Inoue2016}), specifically in initial brightening regions. Using spectropolarimetric data obtained by IBIS to examine the orientation of field lines during the flare, we found that the scenario resembles untwisting of field lines during the flare, as observed by \citet{Kleint2017}. As the flare progresses,  ﬂare loops brighten successively from lower to higher atmospheric layers \citep{Zhang2017} and most of the higher sigmoidal structures continue to exist in their sheared form rather than getting relaxed after the flare (Fig~\ref{fig2}c).


The tether-cutting reconnection in low lying sheared field lines leads to the formation of filaments near the PIL region \citep{Moore2001}. These filaments which are underneath the high lying sigmoids form the sigmoid-filament systems, which undergo apparent merging to form an elongated filamental structure in chromosphere, as observed in GONG H$\alpha$ images, and are co-spatial with an inverse S-shaped sigmoid in the higher layers, as revealed in EUV/AIA observations. Once the filaments merge together to form a long filament, the sigmoid footpoints were found to have temperature and density differences, as shown by DEM analysis. The temperature and density differences between the sigmoid footpoints mostly cause the streaming and counter-streaming of plasma inside the filament. The average flow velocity directed towards the footpoints of the filament is found to be about 40 km s$^{-1}$, in agreement with past studies \citep{Wang1999}, leading to a density decrease by more than 50\%. The continous streaming of chromospheric material of the filament at its footpoints leads to draining of the filament mass (supplement movie). As the total mass of the filament decreases, the sigmoid holding the filament becomes unstable and consequently starts to rise and expands in upward direction \citep{Zhang2021}. However, the sigmoid-filament system could not proceed with its outward motion, but instead it splits axially (Fig~\ref{fig6}). We note that majority of filament eruptions are studied by considering negligible pressure and mass of filament plasma suspended by a ﬂux rope in comparison with the dominant magnetic pressure and tension forces of the flux rope and its surroundings \citep{Titov1999}. However, a few studies \citep{Seaton2011,Jenkins2018}, including this one, provide evidences for “mass-unloading” as an eruption driver or increase the height of flux rope, suggests that a modification of gravitational force due to reduction in mass may influence the stabilization of flux ropes.

The filaments that are formed in between the flare ribbons  along the PIL started to appear in H-alpha images around 22:30 UT (Fig~\ref{fig6}a), only after the flare peak time (i.e., 21:41 UT) but as a result of long flare magnetic reconnection. AIA/SDO observations revealed the stratified structure of flare loops and each set of flare loops did not undergo significant ascending or descending motions after the flare peak time \citep{Zhang2017}, which is corroborated by observations of no considerable lateral separation of flare ribbons \citep{Thalmann2015}. This further substantiates the fact that the same magnetic field structure undergoes reconnection repeatedly for a long period of time, leading to the formation of filaments. Further, the sigmoidal filament structure formed after the flare peak lies along the main PIL with its footpoints rooted at the two flare ribbons on either side of the PIL (animation video), and the dynamics of filament evolution like its rise motion and separation, are all closely related to thermodynamic properties of the same set of flaring loops rooted at flare ribbons, which all get constrained under the same canopy of strong external field within the flare duration of 5-6 hours. Thus, we believe that there is an inherent association of dynamics and non-eruptiveness of filament to the occurrence and confinement nature of X3.1 flare.

Regarding the causes for confinedness of the flare, magnetic reconnection in the low lying, sheared core field is supposed to reduce the constraints of overlying field lines and to allow the core field to erupt \citep{Antiochos1999,Moore2001}. However, in the present event, flare loops did not undergo ascending or descending motions after the flare peak time, suggesting that the tether-cutting reconnection failed to weaken the constraints of upper magnetic loops and to produce the eruption of formed filaments. This is not a new result, for example \citet{Zou2019} studied a confined X2.2 flare which exhibited two episodes of brightenings. They found that these brightenings correspond to two magnetic reconnections, one occurred at the null point beside the pre-existing flux rope and the other tether-cutting reconnection occurred below the flux rope. However, these two magnetic reconnections failed to produce an eruption because of the strong strapping flux. Thus, although tether-cutting reconnection may act as the trigger of an eruption, it alone is less likely to produce a successful eruption.


In eruptive flares, the ejection of twisted flux ropes into interplanetary space leads to less sheared post-flare loops \citep{Forbes1991}. On the other hand, in a confined flare, like the one we have investigated, twist and shear of the core field is conserved with minimum changes in morphological complexity, as shown in Fig~\ref{fig9} \& \ref{fig2}. These are the characteristic effects of confinedness of the flare. Further investigation of non-potentiality of the core AR 12192 suggests that the AR has fragmented and scattered high gradient PILs, which is an indication of not having a continuous, strong twisted flux-rope capable of eruption at certain instability conditions \citep{Vemareddy2019}. This in turn is in agreement with the full neutralization condition (NCN =1) of AR 12192, indicating the absence of a direct current channel over the PIL. On the contrary, sigmoid-filament structure appeared along the main PIL of AR and exhibited dynamics of rise and expansion. Thus, the main contribution to the confinedness of the X3.1 flare should be the stronger inward directed force from background field and not the weaker outward driving force from the magnetic non-potentiality of core AR.

AR 12192 has a mean area of more than 3500 millionths of a solar hemisphere ($\mu$sh) and a peak area of more than 4000 $\mu$sh on 24 October \citep{Cliver2022} with a total unsigned magnetic flux ($|\phi|$) larger than $1\times10^{23}$ Mx. Recent statistical studies \citep{Li2020,Cliver2022} showed that the probability of producing eruptive flares by an AR with area above $\approx$ 3500 $\mu$sh and $|\phi|$ above $1\times10^{23}$ Mx is greatly reduced. They argued that larger the flux and area, stronger will be the confinement of the overlying magnetic field. This argument holds true even for the location of the X3.1 flare, which occurs near the center of the AR. Statistically, \citet{Baumgartner2018} showed that confined flares occur close to the AR centers, where the constraining field strength is stronger and eruptive flares occur at the periphery of ARs, where the confinement is weaker. The total flux and area of AR along with the location of the X3.1 flare indicate the increase of horizontal field strength, which decreases the decay rate of overlying field with height, suppressing eruption. The average critical height (height at which the decay index = 1.5) above the sigmoid-filament system remains constant at about $\approx$ 80 Mm throughout the flare duration of 5-6 hours, suggesting the strong confinement over the core of the AR \citep{Vasantharaju2018,Baumgartner2018,Li2020}. 

It is very difficult to point out the exact component of downward Lorentz force, generated from the interaction between external field and the erupting structure, contributing towards confining the eruption with pure observations. Given the fact that the X3.1 flare event is an on-disk event and the non-availability of  STEREO observations, it is difficult to determine the exact height till which the merged filament raised before it actually got suppressed by downward acting Lorentz force. However, based on AIA 171~\AA~and 304~\AA~observations (low temperature channels), the filament eruption is confined in the lower corona ($<$ 80 Mm) and the decay index of the external poloidal ﬁeld does not exceed the criterion for torus instability (i.e. when $n_c$ = 1.5, $H_c$ = 80 Mm). Further, the potential field configuration at different time instants throughout the flare duration (Fig~\ref{fig10}) provide evidences that the direction of the external poloidal magnetic field is oriented nearly perpendicular to the axial direction of filament (along the PIL). This indicates that the strapping force is a more dominant  downward force than the dynamic tension force, induced by the toroidal field \citep{Myers2015}. However, we can not rule out the possibility of downward acting non-axisymmetry induced forces due to the radial magnetic field of the magnetic flux rope carrying the filament \citep{Zhong2021}. The direction of the forces induced by the radial magnetic field of the filament changes with the evolution of the filament but determining them using the observations is very hard. So, we conclude that from an observational point of view the confinedness of the X3.1 event is due to the net downward Lorentz force contributed mainly by the strapping field with the possible contribution from non-axisymmetry of the filament.

More of such unique X-class confined events need to be analyzed to generalize the results reported in this work and to provide reliable input to flare/CME forecasting studies.

\section{Acknowledgements}	
We thank the referee for the detailed comments that definitely improved the quality of the paper. N.V. acknowledges support from the European Union’s Horizon 2020 research and innovation programme under grant agreement no. 824135 (SOLARNET project) and no. 739500 (PRE-EST project). This research has been carried out in the framework of the CAESAR (Comprehensive spAce wEather Studies for the ASPIS prototype Realization) project. N.V. acknowledges financial contribution from the Agreement ASI-INAF n.2020-35-HH.0. This work was also supported by the Italian MIUR-PRIN grant 2017APKP7T, by the Università degli Studi di Catania (Piano per la Ricerca Università di Catania 2020-2022, Linea di intervento 2). SDO is a mission of NASA's Living With a Star Program. Authors thank the HMI and AIA science team for their open data policy. We also thank Dr. Christian Beck for providing the IBIS calibrated data. H$\alpha$ data was acquired by GONG instruments operated by NISP/NSO/AURA/NSF with contribution from NOAA. Hinode is a Japanese mission developed and launched by ISAS/JAXA, with NAOJ as domestic partner and NASA and STFC (UK) as international partners. N.V. acknowledges Marianna B. Korsós for providing usage instructions of ParaView visualization software.



\bibliographystyle{apj}
\bibliography{12192_ref}

\end{document}